\documentclass[a4paper,fleqn,10pt]{article}
\pdfoutput=1
\usepackage{amsmath}
\usepackage{amssymb}
\usepackage{array}
\usepackage{calc}
\usepackage{longtable}
\usepackage{multirow,booktabs}
\usepackage{relsize}
\usepackage{pstricks}
\usepackage{graphicx}
\usepackage{xspace}
\usepackage{units}

\numberwithin{equation}{section}
\usepackage{mciteplus}
\usepackage[a4paper,pdfborder={0 0 0}]{hyperref}
\usepackage[format=hang,labelfont=bf,hypcap=true]{caption}
\usepackage{subcaption}
\usepackage{sectsty}
\allsectionsfont{\sffamily}
\subsubsectionfont{\mdseries\itshape\large}
\makeatletter
\DeclareRobustCommand*{\bfseries}{%
  \not@math@alphabet\bfseries\mathbf
  \fontseries\bfdefault\selectfont
  \boldmath
}
\makeatother
\let\spreprint\empty
\newcommand{\preprint}[1]{\def\spreprint{\protect#1}}
\let\sinstitute\empty
\newcommand{\institute}[1]{\def\sinstitute{\protect#1}}
\makeatletter
\renewcommand{\maketitle}{\begingroup
  \null\thispagestyle{empty}%
    \ifx\spreprint\empty
      \vskip 5ex
    \else
      \flushright\large\spreprint\vskip 2ex
    \fi
    \vskip 5ex
    \flushleft
      {\sffamily\bfseries\huge\@title}\vskip 6ex
      \@author\vskip 2ex
      \ifx\sinstitute\empty
      \else
        {\small\sinstitute}
      \fi
    \vskip 5ex
  \endgroup
}
\makeatother
\renewenvironment{abstract}{\begin{center}
  {\large\sffamily\bfseries Abstract: }
  \begin{minipage}[t]{0.75\textwidth}
}{\end{minipage}\end{center}\vskip 10ex}

\numberwithin{equation}{section}
\DeclareRobustCommand{\plusplus}{\raisebox{0.2ex}{\smaller++}}

\newcommand{\bea}{\begin{align}}
\newcommand{\eea}{\end{align}}
\newcommand{\beq}{\begin{equation}}
\newcommand{\eeq}{\end{equation}}

\newcommand{\bs}{\begin{split}}
\newcommand{\es}{\end{split}}

\newcommand{\bi}{\begin{itemize}}
\newcommand{\ei}{\end{itemize}}
\newcommand{\bc}{\begin{center}}
\newcommand{\ec}{\end{center}}
\newcommand{\bac}{\begin{array}{c}}
\newcommand{\bacc}{\begin{array}{cc}}
\newcommand{\ea}{\end{array}}

\def\spa#1.#2{\langle#1\,#2\rangle}
\def\spb#1.#2{[#1\,#2]}

\newcommand{\UGeV}{\ensuremath{\mathrm{GeV}}\xspace}

\newcommand{\MZ}{\ensuremath{m_Z}}

\newcommand{\alphaS}{\ensuremath{\alpha_\text{s}}\xspace}

\newcommand{\Qcut}{\ensuremath{Q_\mathrm{cut}}}

\newcommand{\sla}[1]{\ensuremath{{#1\kern-0.45em/}}}

\newcommand\LHC{L\protect\scalebox{0.8}{HC}\xspace}
\newcommand\ATLAS{\atlas}
\newcommand\atlas{A\protect\scalebox{0.8}{TLAS}\xspace}
\newcommand\CMS{\cms}
\newcommand\cms{C\protect\scalebox{0.8}{MS}\xspace}

\newcommand{\MCatNLO}{M\protect\scalebox{0.8}{C}@N\protect\scalebox{0.8}{LO}\xspace}

\newcommand{\MEPSatLO}{M\protect\scalebox{0.8}{E}P\protect\scalebox{0.8}{S}@L\protect\scalebox{0.8}{O}\xspace}
\newcommand{\MEPSatNLO}{M\protect\scalebox{0.8}{E}P\protect\scalebox{0.8}{S}@N\protect\scalebox{0.8}{LO}\xspace}

\newcommand{\Gosam}{G\protect\scalebox{0.8}{O}S\protect\scalebox{0.8}{AM}\xspace}

\newcommand{\Collier}{C\protect\scalebox{0.8}{OLLIER}\xspace}

\newcommand{\BlackHat}{B\protect\scalebox{0.8}{LACK}H\protect\scalebox{0.8}{AT}\xspace}

\newcommand{\OpenLoops}{O\protect\scalebox{0.8}{PEN}L\protect\scalebox{0.8}{OOPS}\xspace}

\newcommand{\Njet}{N\protect\scalebox{0.8}{JET}\xspace}

\newcommand{\Sherpa}{S\protect\scalebox{0.8}{HERPA}\xspace}
\newcommand{\Comix}{C\protect\scalebox{0.8}{OMIX}\xspace}

\newcommand{\Amegic}{A\protect\scalebox{0.8}{MEGIC\plusplus}\xspace}

\newcommand{\Rivet}{R\protect\scalebox{0.8}{IVET}\xspace}

\newcommand{\Fastjet}{F\protect\scalebox{0.8}{AST}J\protect\scalebox{0.8}{ET}\xspace}

\hypersetup{
  pdfauthor={}
  pdftitle={}
}
\preprint{IPPP/16/121\\DCPT/16/242\\MCNET-16-46}
\author{Frank Krauss$^1$, Davide Napoletano$^1$, Steffen Schumann$^2$}
\title{Simulating $b$-associated production of $Z$ and Higgs bosons with \Sherpa}
\institute{
  $^1$ Institute for Particle Physics Phenomenology,
  Durham University, Durham DH1 3LE, UK\\
  $^2$ II. Physikalisches Institut, Georg-August-Universit\"at G\"ottingen, 37077 G\"ottingen, Germany}
\begin{document}
\maketitle
\begin{abstract}
  We compare four-- and five--flavour scheme predictions for $b$-associated
  production of $Z$ and Higgs bosons. The results are obtained with \Sherpa's
  \MCatNLO implementation for the four--flavour scheme, treating the $b$'s
  as massive, and with multijet merging at leading and next-to leading
  order for the five--flavour schemes. Comparison with data for $Z+b(\bar{b})$
  production at the $7$ TeV LHC exhibit strengths and weaknesses of the
  different approaches and are used to validate predictions for $b$-associated
  Higgs-boson production at the 13 TeV Run II.
\end{abstract}
\section{Introduction} 

The production of $Z$ and $H$ bosons in association with $b$-quarks or 
through $b\bar{b}$-\-annihilation has recently attracted renewed interest,
for a number of old and new reasons~\cite{
  Frederix:2011qg, Wiesemann:2014ioa,deFlorian:2016spz,Lim:2016wjo}.

Firstly, the associated production of a vector boson ($V$) and a $b$-tagged 
jet points to underlying processes like $gb\to Vb$ at Born--\-level and is
thus sensitive to the $b$-quark parton distribution function (PDF).
These are particularly important for phenomenologically relevant processes
as the production of a Higgs boson in $b\bar{b}$--\-annihilation, in 
association with $b$ jets or a single top.  The latter processes contribute 
to the total Higgs-boson production in the Standard Model on the level of a 
few percent and must therefore be included in fits to the couplings of the 
Higgs boson~\cite{deFlorian:2016spz}.

Secondly, and complementary to this purely Standard-Model reasoning, many
models for physics beyond the Standard Model come with extended Higgs sectors,
quite often in the form of a second Higgs doublet.  The mixing among the
Higgs doublets often amplifies the couplings of the Higgs bosons to the
$b$-quarks.  As a consequence, $bH$ and $b\bar{b}H$ production provide
important search grounds for new physics. Furthermore, events with identified
$b$-jets and a significant missing transverse momentum constitute a possible
signature of Dark Matter~\cite{Lin:2013sca,Aad:2014vea,Chen:2016ylf} production. 
For this signal invisibly decaying $Z$-bosons associated with $b$-jets 
pose a severe irreducible background. 

In addition to the processes considered in this work also hadronic single
top-quark production through the t-channel process proceeds via initial-state
$b$-quarks.  At leading order this corresponds to the process $qb\to tq'$.
A proper treatment of the initial-state $b$-quark and of higher-order
processes such as $qg\to tq'\bar{b}$ is again vital for the successful
description of this important signature. 

In all cases, however, the treatment of the $b$-quark is far from being 
straightforward, since commonly PDFs are assumed to be valid for massless
partons only -- parton masses induce logarithmic and unknown power corrections.
On the other hand the $b$-quark mass is large enough, around
4.5 GeV, to induce visible kinematic effects for jets.   With a jet transverse
momentum of about 20 GeV they can be estimated by $(m_b/p_T)^2$, on the level
of around 10\%.  Aiming for such accuracies, the treatment of the $b$-quark
mass therefore poses a problem.  One solution is the five--flavour
scheme~\cite{Harlander:2003ai}, defined by assuming the $b$-quark as strictly
massless in the matrix elements and by allowing a non-vanishing $b$-quark PDF.
In the context of our studies this translates into using multijet-merging
technology to combine processes such as $b\bar{b}\to H$, $gb\to Hb$,
$gg\to Hb\bar{b}$ etc.\  with $m_b=0$ into a fully inclusive sample.
On the other hand, from an alternative point of view, one could also claim
that $b$-quark PDFs were ill-defined objects and would therefore set them
to zero.  In this case, one would study processes such as $gg\to Zb\bar{b}$ and 
$gg\to Hb\bar{b}$~\cite{Dawson:2003zu,FebresCordero:2006sj} instead, 
and use the finite $b$-quark mass to regularise the otherwise divergent
phase-space integrals.  This treatment, namely taking the $b$-quark mass
fully into account but releasing any phase-space constraints on their final
state, defines the four--flavour scheme. 

Various ways of combining results obtained in these two schemes have been 
proposed.  Among them, the so-called {\it Santander-Matching} first presented 
in~\cite{Harlander:2011aa} is probably the most widely used one.  Its
approach is to combine four-- and five--\-flavour scheme predictions by
means of a dynamically weighted average of them.  This weight is defined to
be a continuous function of the hard scale of the process and the mass of
the bottom quark, in such a way that when the ratio between the hard scale
and the $b$-quark mass is large, the five--\-flavour prediction is recovered,
while for scales of similar size the four--\-flavour result is obtained.

As a well-defined alternative, the FONLL approach has been introduced for
$b$-quark hadro-production~\cite{Cacciari:1998it}.  It has later been
extended to deep--\-inelastic scattering (DIS)~\cite{Forte:2010ta} and
lately it has been applied to Higgs production in bottom quark
fusion~\cite{Forte:2015hba,Forte:2016sja}.  The main idea of this method is
to take the four-- and five--\-flavour scheme perturbative-series expansions
and, after having re-arranged them in such a way that they become compatible,
to isolate double-counting terms in the two schemes. This is achieved by
re-expressing PDF evolution and the running of $\alpha_S$ in the
four--\-flavour scheme in terms of those computed in the five--\-flavour
scheme, and heavy-flavour PDFs in the five--\-flavour scheme in terms of
light-flavour PDFs.  The final prediction is obtained by replacing terms
in the five--\-flavour scheme by their known counterparts computed in the
four--\-flavour scheme. 

Another matching procedure, based on an Effective Field Theory approach,
has recently been developed in Refs.~\cite{Bonvini:2015pxa, Bonvini:2016fgf}.
More methods to match initial-state resummed massless predictions with
fixed-order ones are available in DIS physics.  In general, decoupling
schemes, like the four--\-flavour scheme, in this context, are referred to as 
Fixed--\-Flavour--\-Number--\-Schemes (FFNS) while massless schemes, in which 
the number of active flavours changes with energy, like the five--\-flavour 
scheme, are called Variable--\-Flavour--\-Number--\-Schemes (VFNS).  Adding 
mass effects to a VFNS leads to a General-Mass(GM)-VFNS (as opposed to the 
completely massless Zero--\-Mass(ZM)--\-VFNS).  Both Santander and the FONLL 
matching together with the ACOT scheme in different 
versions~\cite{Aivazis:1993pi,Aivazis:1993kh,Kramer:2000hn} fall into the 
GM-VFNS class.

This discussion and the corresponding schemes apply mainly to analytic
calculations.  As soon as fragmentation effects are to be accounted for, the
finite $b$-quark mass must be taken into account, as it reduces the emission
rate of gluons off the quark with respect to the strictly massless case.
To avoid a resulting fragmentation function which would be significantly
too soft, $b$-quarks need to be treated as massive within parton-shower
simulations.  Contact with massless calculations is established by
shifts of the four-momenta of $b$-quarks before or during their first
emission.

The simulation of final states with variable jet multiplicities, i.e.\
varying levels of inclusiveness, is the realm of matrix-element
parton-shower matching and merging techniques~\cite{Buckley:2011ms}. They
combine the strengths of both approaches.  Exact leading-order (LO) or
next-to-leading-order (NLO) QCD matrix elements describe hard, well-separated
parton configurations, while additional softer jets and in general jet
evolution is accounted for by parton showers. Considering final states with
{\em identified} $b$-jets certainly poses stringent tests on these algorithms
and in fact requires dedicated methods to correctly account for the
non-vanishing $b$-quark mass in both ingredients of the calculations. 

In this publication we discuss and validate the corresponding methods
within the \Sherpa event generator~\cite{Gleisberg:2003xi,Gleisberg:2008ta}.  
Different choices for treating the $b$-quarks in the matrix elements, 
ranging from massless in a five--\-flavour scheme (5FS) to massive in a 
four--\-flavour scheme (4FS), consistently combined with parton showers, 
are compared.  The presented approaches are implemented and readily available
from \Sherpa-2.2.1. 

The discussion is organised as follows.  In Sec.~2, the underlying calculations
are briefly reviewed.  In addition, the different flavour schemes 
are defined in more detail.  In Sec.~3, using $Z+b$ and $Z+b\bar{b}$ data at 
$7$ TeV from Run I of the \LHC, the methods are validated, and their relative 
strengths and weaknesses are identified.  The findings will be used in the next 
section, Sec.~4, to arrive at informed and robust predictions for the 
$b$-associated production of Higgs bosons at the 13 TeV Run II of the \LHC.  
In the conclusions some comments concerning further implications for BSM 
physics put this study into a wider context.

\section{Calculational Methods and Setups}
\label{sec:setup}

Efficient routines for the required QCD matrix-element calculations and a
well understood QCD parton-shower are the key ingredients to all matching
and merging calculations.  Within \Sherpa LO matrix elements are
provided by the built-in generators \Amegic~\cite{Krauss:2001iv} and
\Comix~\cite{Gleisberg:2008fv}.  While virtual matrix elements contributing to 
QCD NLO corrections can be invoked through interfaces to a number of 
specialised tools, e.g.\ \BlackHat~\cite{Berger:2008sj}, \Gosam~\cite{Cullen:2014yla}, 
\Njet~\cite{Badger:2012pg}, \OpenLoops~\cite{Cascioli:2011va} or through
the BLHA interface~\cite{Binoth:2010xt}, we employ in this study the
\OpenLoops generator~\cite{OL_hepforge} in conjunction with the \Collier 
library~\cite{Denner:2016kdg,Denner:2014gla}.  Infrared divergences are
treated by the Catani--Seymour dipole method~\cite{Catani:1996vz,Catani:2002hc}
which has been automated in \Sherpa~\cite{Gleisberg:2007md}.   In this
implementation mass effects are included for final-state splitter and
spectator partons but massless initial-state particles are assumed throughout.
\Sherpa's default parton-shower model~\cite{Schumann:2007mg,Hoeche:2009xc}
is based on Catani--Seymour factorisation~\cite{Nagy:2006kb}.  In order to
arrive at meaningful fragmentation functions for heavy quarks, all modern
parton showers take full account of their finite masses in the final state,
although in algorithmically different ways.  In \Sherpa, the transition from
massless to massive kinematics is achieved by rescaling four-momenta at the
beginning of the parton shower.  In the initial-state parton shower in \Sherpa,
the $g\to b\bar{b}$ and $b\to bg$ splitting functions do not contain $b$-quark
mass effects in their functional form and account for mass effects in the
kinematics only.

In the following we briefly define the methods available in \Sherpa
for simulating $b$-associated production processes, that will then be
validated and applied for LHC predictions:
\begin{description}
\item[4F NLO (4F \MCatNLO):] In the {\em four--flavour scheme}, $b$-quarks
  are consistently treated as {\em massive} particles, only appearing in
  the final state.  As a consequence, $b$-associated $Z$- and $H$-boson
  production proceeds through the parton-level processes $gg\to Z/H+b\bar{b}$,
  and $q\bar{q}\to Z/H+b\bar{b}$ at Born level.  \MCatNLO matching is
  obtained by consistently combining fully differential NLO QCD calculations
  with the parton shower, cf.~\cite{Frixione:2002ik,Hoeche:2011fd}.  Due to
  the finite $b$-quark mass these processes do not exhibit infrared 
  divergences and the corresponding {\em inclusive} cross sections can thus 
  be evaluated without any cuts on the $b$-partons.
\item[5F LO (5F \MEPSatLO):] In the {\em five--flavour scheme} $b$-quarks
  are {\em massless} particles in the {\em hard matrix element}, while they
  are treated as massive particles in both the initial- and final-state
  {\em parton shower}.  

  In the \MEPSatLO~\cite{Hoeche:2009rj} samples we merge $pp \rightarrow H/Z$
  plus up to three jets at leading order; this includes, for instance, the
  parton--level processes $b\bar{b} \to Z/H$, $gb\to Z/H b$,
  $gg\to Z/H b\bar{b}$, $\dots$.  To separate the various matrix-element
  multiplicities, independent of the jet flavour, a jet cut of
  $\Qcut = 10\,\UGeV$ is used in the $Z$ case while $\Qcut = 20\,\UGeV$ is
  employed in $H$-boson production.
\item[5F NLO (5F \MEPSatNLO):] In the 5FS \MEPSatNLO
  scheme~\cite{Gehrmann:2012yg,Hoeche:2012yf}, we account for quark masses in
  complete analogy to the LO case: the quarks are treated as massless in the
  hard matrix elements, but as massive in the initia- and final-state parton
  showering.  Again, partonic processes of different multiplicity are merged
  similarly to the \MEPSatLO albeit retaining their next-to-leading-order
  accuracy.  In particular, we consider the merging of the processes
  $pp\rightarrow H/Z$ plus up to two jets each calculated with \MCatNLO
  accuracy further merged with $pp\rightarrow H/Z + 3j$ calculated at
  \MEPSatLO.
\end{description}
  
We consistently use four--flavour PDFs in the 4F
scheme, i.e.\ the dedicated four--flavour NNPDF3.0 set~\cite{Ball:2014uwa} with
the strong coupling given by $\alphaS(\MZ)\,=\,0.118$ and running at NLO.
For the simulations in the five--flavour schemes the five--flavour NNLO PDFs
from NNPDF3.0 are used, with $\alphaS(\MZ)\,=\,0.118$ and running at NNLO.  
We assume all quarks apart from the $b$ to be massless, with a pole mass of
$m_b = 4.92$~\UGeV which enters the hard matrix-element calculation, where
appropriate, and the parton shower.

Results in the 4F and 5F schemes have been obtained with the default scale-setting 
prescription for parton-shower matched calculations in \Sherpa~\cite{Hoeche:2009rj,Hoeche:2010av}.
They are 
calculated using a backward-clustering algorithm, and for each emission from the shower, couplings 
are evaluated at either the $k_T$ of the corresponding emitted particle (in the case of
gluon emission), or at the invariant mass of the emitted pair (in the case
of gluon splitting into quarks).  The clustering stops at a ``core'' $2\to 2$
process, with all scales set to $\mu_F=\mu_R=\mu_Q=m_T(V)/2$, where $m_T(V)$
corresponds to the transverse mass of the boson. This scale is thus used to evaluate
couplings in the hard matrix element and PDFs.
The corresponding central values are supplemented with uncertainty bands
reflecting the dependence on the unphysical scales.  Renormalisation and
factorisation scales are varied around their central value by a factor of
two up and down, with a standard 7-point variation.  The scale variations use 
the \Sherpa internal reweighting procedure~\cite{Bothmann:2016nao} and result in
envelopes around the central value. Furthermore, we consider explicite variations 
of the parton-shower starting scale, i.e. $\mu_Q$, by a factor of two up and down. 

For the Higgs-bosons production processes, bottom and top Yukawa couplings are
important.  Using their corresponding pole mass, $m_b = 4.92$~\UGeV and
$m_t = 172.5$~\UGeV, and, subject to a LO running they are finally evaluated
at $\mu_{m} = m_H=125$~\UGeV. We do not include variations of this scale.

\section{Bottom-jet associated Z-boson production}
\label{sec:zbb}
The production of a $Z$ boson in association with QCD jets provides the 
ideal test bed for the theoretical approaches outlined above.  Through 
the decay of the $Z$ boson to leptons these processes yield a rather 
simple and clean signature with sizeable rates even for higher
jet counts. Precise measurements of the production rates and differential 
distributions of both the $Z$-boson decay products and the accompanying 
jets offer discriminating power for miscellaneous theoretical approaches.
In fact, measurements of $Z+$jets production served as key inputs for
the validation of matrix-element parton-shower simulation techniques,
cf.~\cite{Aad:2013ysa,Khachatryan:2014zya,Khachatryan:2016crw}, 
and  impressively underpin the enormous success of these calculational
methods.

Here we focus on the production of $Z$ bosons accompanied by identified
$b$-jets. Comparison with data from both the ATLAS and CMS collaborations at
$7$ TeV~\cite{Aad:2014dvb,Chatrchyan:2013zja} provides the benchmark for the 
accuracy and quality of four-- and five--flavour simulations with \Sherpa.
Similar measurements at $8$ and $13$ TeV \LHC collision energies are 
under way~\cite{Khachatryan:2016iob}.

\subsection{Measurements at \LHC Run I -- the reference data}

Based on a data set of $4.6\;{\rm fb}^{-1}$ integrated luminosity the ATLAS
collaboration studied the production of $b$-jets associated with $Z/\gamma^*$
that decay to electrons or muons~\cite{Aad:2014dvb}.  The dilepton 
invariant mass ranges between $76\;{\text{GeV}} < m_{\ell\ell} < 106\;{\text{GeV}}$.  
Jets are reconstructed using the anti-$k_t$ algorithm \cite{Cacciari:2008gp} 
with a radius parameter of $R=0.4$, a minimal transverse momentum of 
$p_{T,j}> 20$ GeV and a rapidity of $|y_{j}|<2.4$.  Furthermore, each jet 
candidate needs to be separated from the leptons by $\Delta R_{j\ell} > 0.5$. 
Jets containing $b$-hadrons are identified using a multi-variate technique. 
To match the outcome of the experimental analysis, simulated jets are 
identified as $b$-jets, when there is one or more weakly decaying $b$-hadron 
with $p_T>5$ GeV within a cone of $\Delta R=0.3$ around the jet axis. The 
sample of selected events is further subdivided into a class containing events 
with at least one $b$-jet ($1$-tag) and a class with at least two $b$-jets 
($2$-tag).

A similar analysis was performed by CMS~\cite{Chatrchyan:2013zja}.  There,
electrons and muons are required to have a transverse momentum of
$p_{T,\ell}>20$ GeV, a pseudorapidity $|\eta_{\ell}|<2.4$, and a dilepton
invariant mass within $81\;{\text{GeV}} < m_{\ell\ell} < 101\;{\text{GeV}}$.
Only events with exactly two additional $b$-hadrons were selected.  The
analysis focuses on the measurement of angular correlations amongst the
$b$-hadrons and with respect to the $Z$ boson. This includes in particular
variables sensitive to rather collinear $b$-hadron pairs. In addition, the 
total production cross section as a function of the vector boson's 
transverse momentum was measured. 

Both analyses are implemented and publicly available in the \Rivet
analysis software~\cite{Buckley:2010ar} that, together with the \Fastjet
package~\cite{Cacciari:2011ma}, is employed for all particle, i.e.\ hadron,
level analyses in this publication. 

\subsection{Comparison with LHC data}

\begin{figure}[!htb]
  \centering{
    \includegraphics[width=0.45\textwidth]{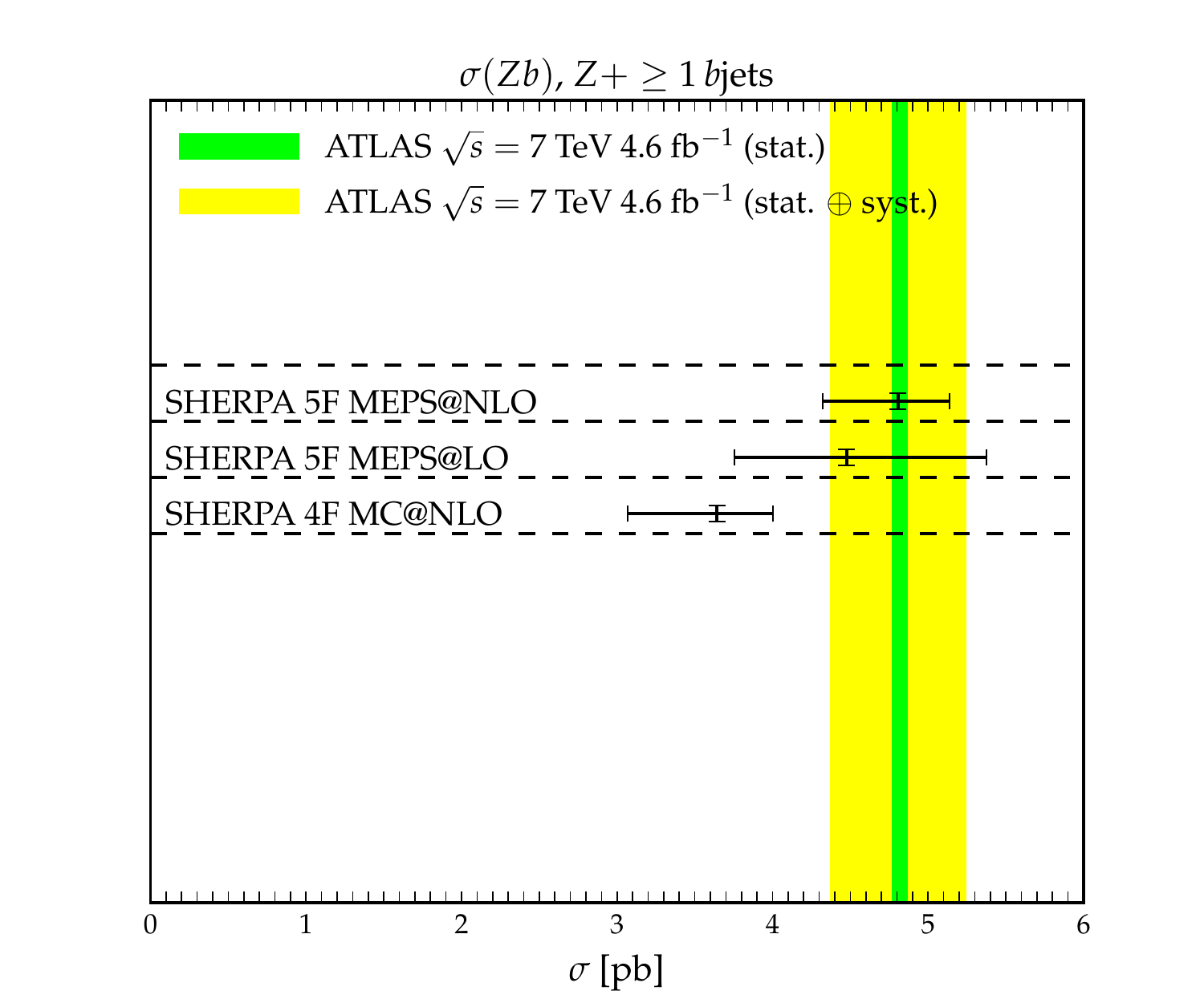}
    \includegraphics[width=0.45\textwidth]{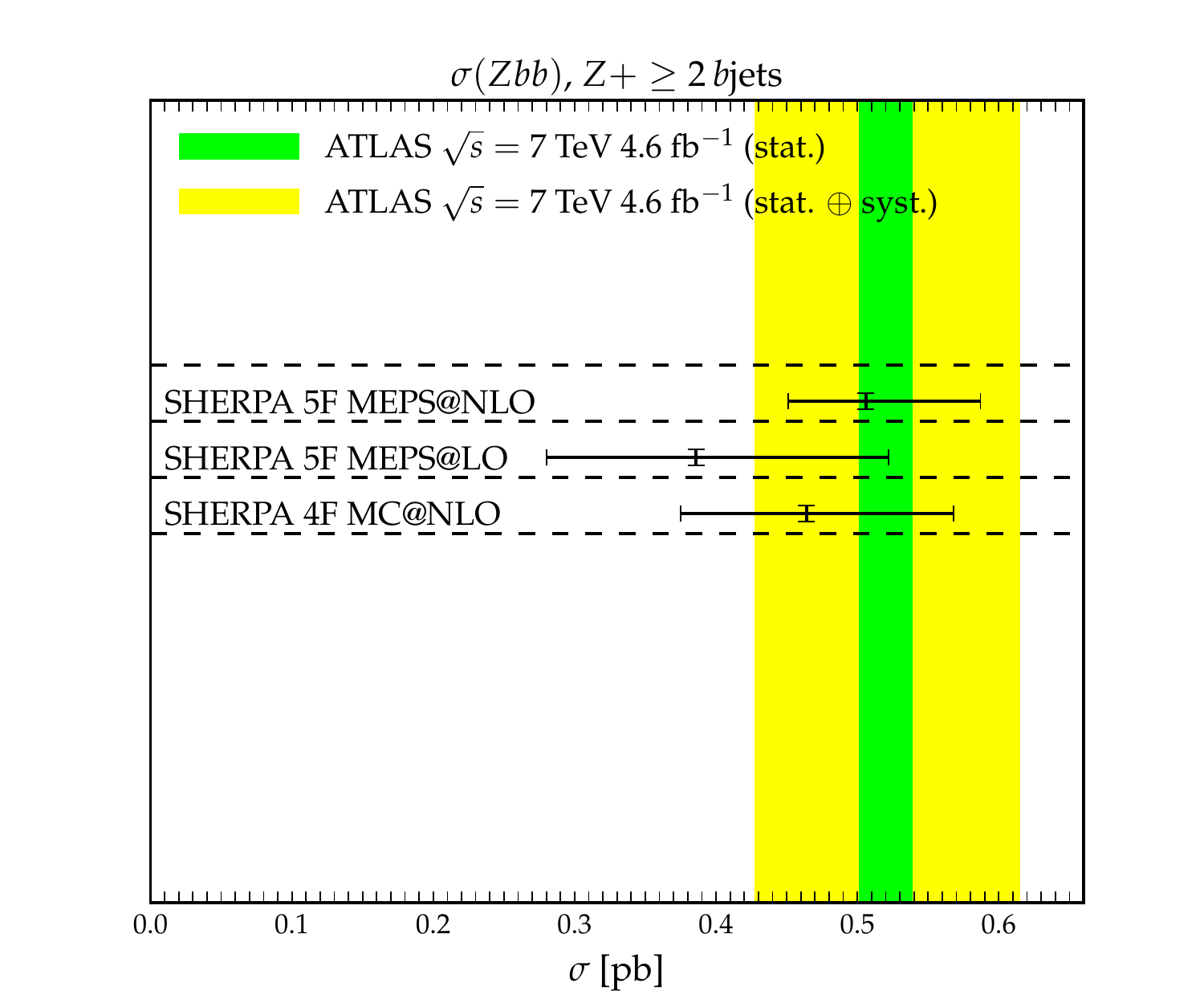}
    \caption{Comparison of total production cross section predictions
      with ATLAS data~\cite{Aad:2014dvb}. The error bars on the
      theoretical results are calculated from variations of the 
      hard-process scales $\mu_{R/F}$ and the parton-shower starting 
      scale $\mu_Q$.
      \label{fig:xstot}}
  }
\end{figure}

In this section the theoretical predictions from \Sherpa will be compared
to the experimental measurements from \LHC Run I.  We begin the discussion
with the comparison with the measurements presented by the ATLAS collaboration
in Ref.~\cite{Aad:2014dvb}.  The total cross sections for $Z+\ge 1$ and
$Z+\ge 2$ $b$ jets are collected in Fig.~\ref{fig:xstot}.  Already there we
see a pattern emerging that will further establish itself in the differential
cross sections: While the 5F \MEPSatNLO results agree very well with data, the
central values of the 5F \MEPSatLO cross sections tend to be around 10-20\%
lower than the central values of data, but with theory uncertainties clearly
overlapping them. For all the runs the uncertainty estimates include both, 
7-point variations of the perturbative scales $\mu_{R/F}$, as well as $\mu_Q$ 
variations by a factor of two up and down. In contrast to the 5F case, the 4F 
\MCatNLO cross sections tend to be significantly below the experimental values 
for the $Z+\ge 1$ $b$-jets cross section, without overlap of uncertainties.  
In the $Z+\ge 2$ $b$-jets cross section the agreement between 4F \MCatNLO 
results and data is better, with the theoretical uncertainties including 
the central value of the measured cross section.  

\begin{figure}[!htb]
\centering{
  \includegraphics[width=0.45\textwidth]{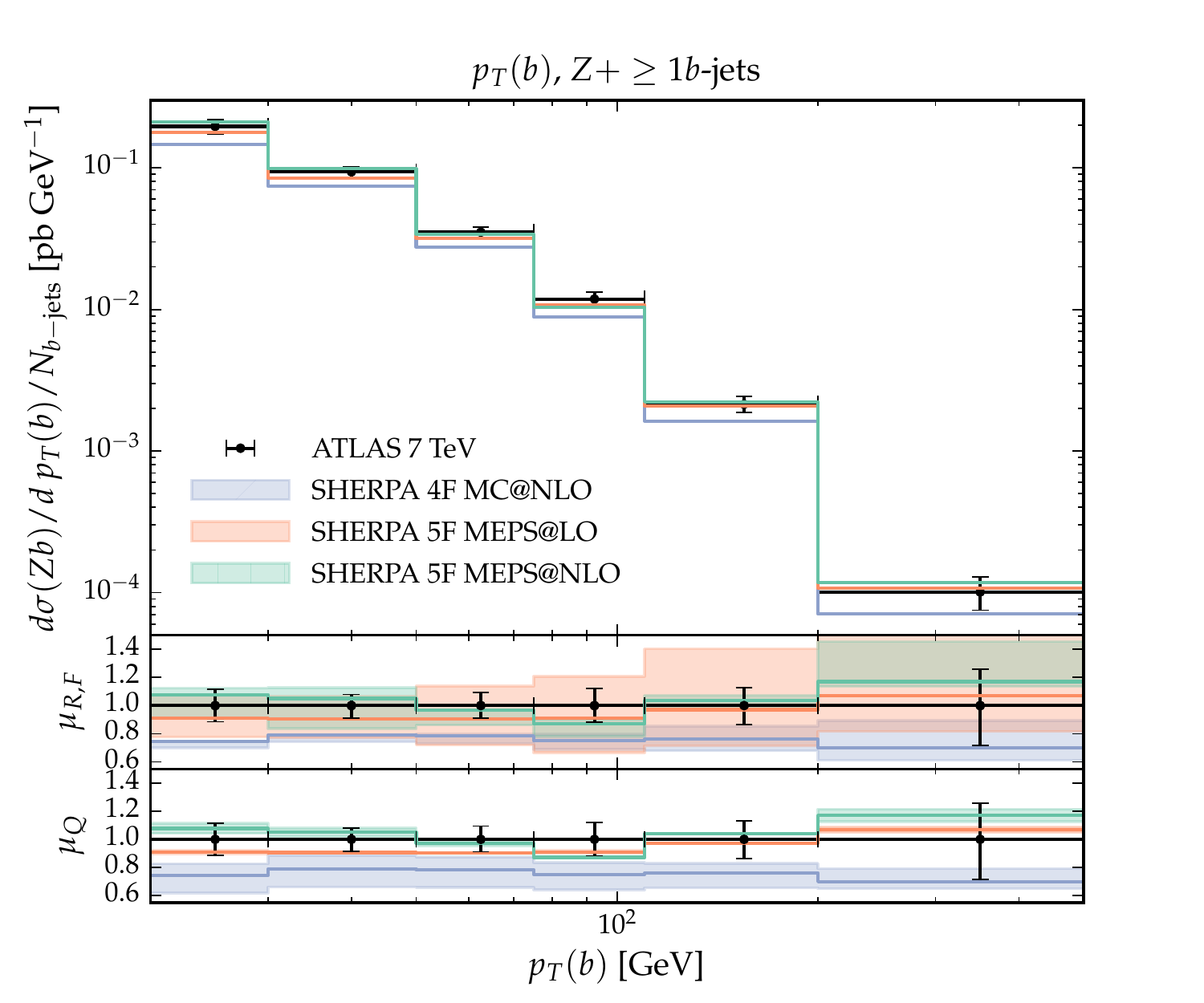}
  \includegraphics[width=0.45\textwidth]{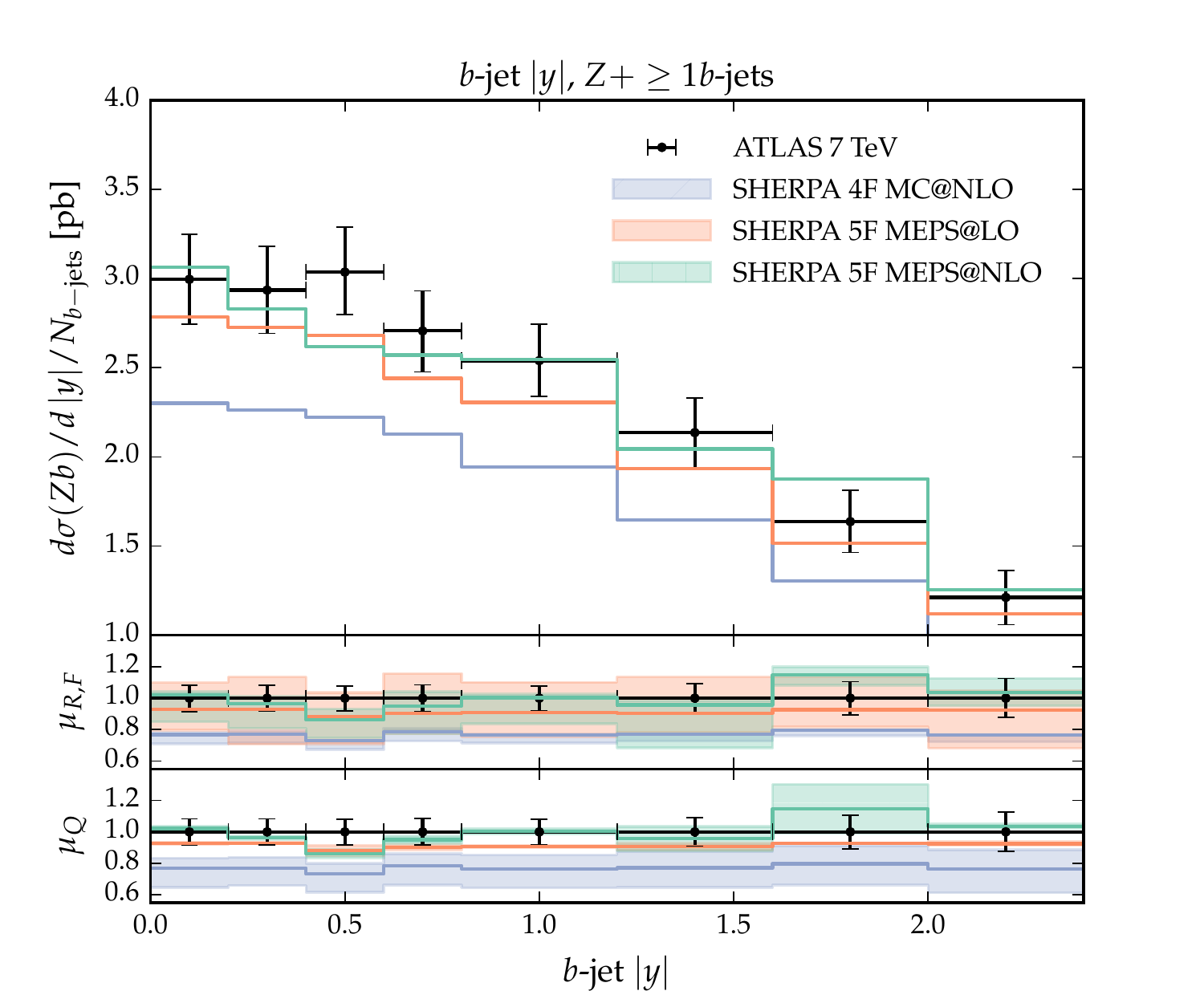}
  \caption{Inclusive transverse-momentum and rapidity distribution of
    all $b$-jets in events with at least one $b$-jet. Data taken
    from Ref.~\cite{Aad:2014dvb}.}
  \label{fig:1b}}
\end{figure}

In Fig.~\ref{fig:1b} the differential cross sections with respect to the
transverse momentum and rapidity of the $b$-jets, normalised to the number of
$b$-jets, are presented for events with {\em at least} one $b$-tagged jet.
The shapes of both distributions are well modelled both by the 4F and
the two 5F calculations.  However, clear differences in the predicted
production cross sections are observed. While the 5F~NLO results are in
excellent agreement with data - both in shape and normalisation - the
central values of the 5F~LO cross sections tend to be around 10\% below
data, at the lower edge of the data uncertainty bands, and
the 4F results are consistently outside data, about 25\% too low.
In the lower panels of Fig.~\ref{fig:1b} and all the following plots
in this section we show the uncertainty bands of the theoretical 
predictions, corresponding to the above described $\mu_{R/F}$ and $\mu_Q$
variations. For the 5FS calculations the scale uncertainties clearly 
dominate, while for the 4F \MCatNLO scheme the shower-resummation 
uncertainty dominates.

\begin{figure}[!tbh]
\centering{
  \includegraphics[width=0.45\textwidth]{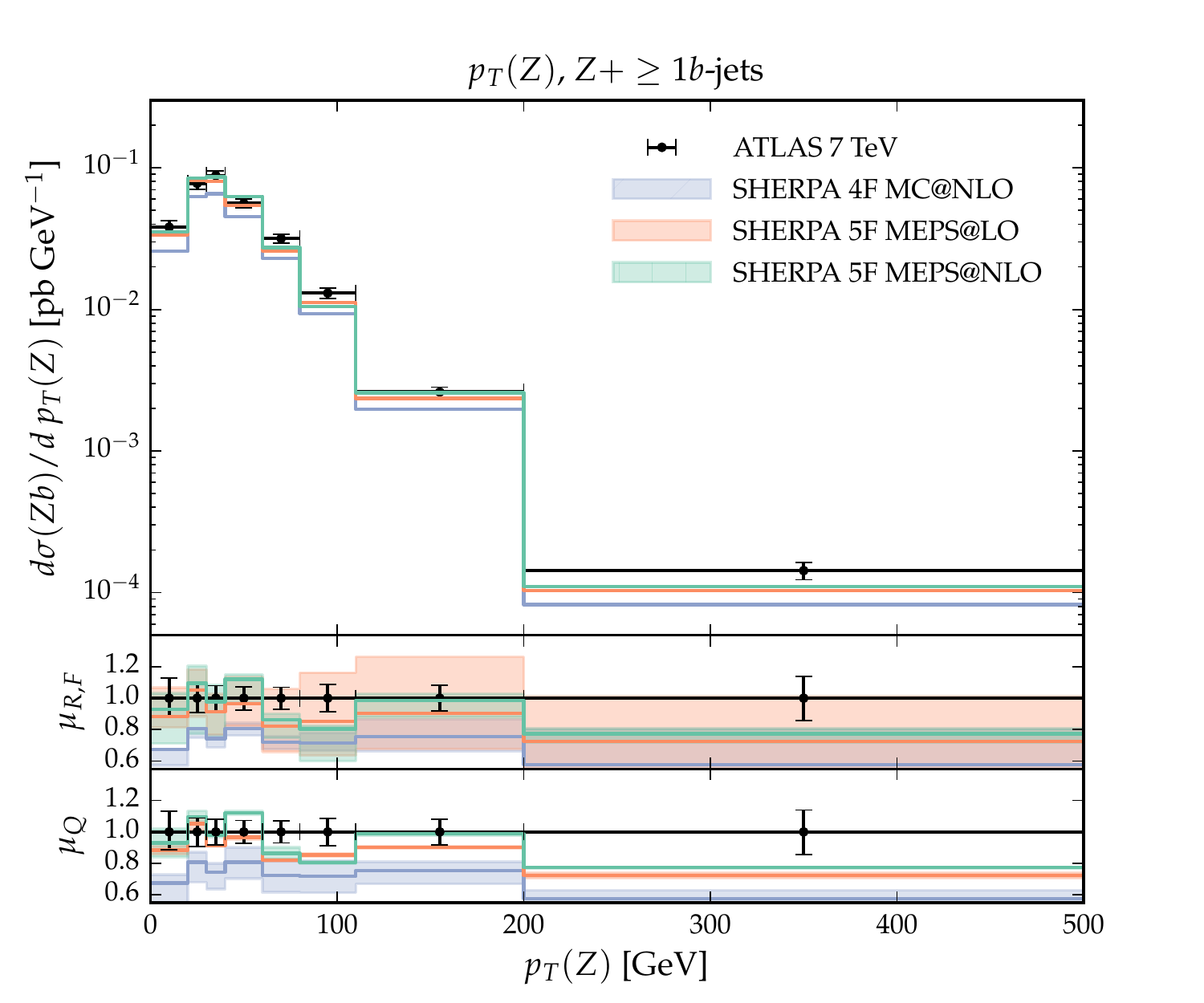}
  \includegraphics[width=0.45\textwidth]{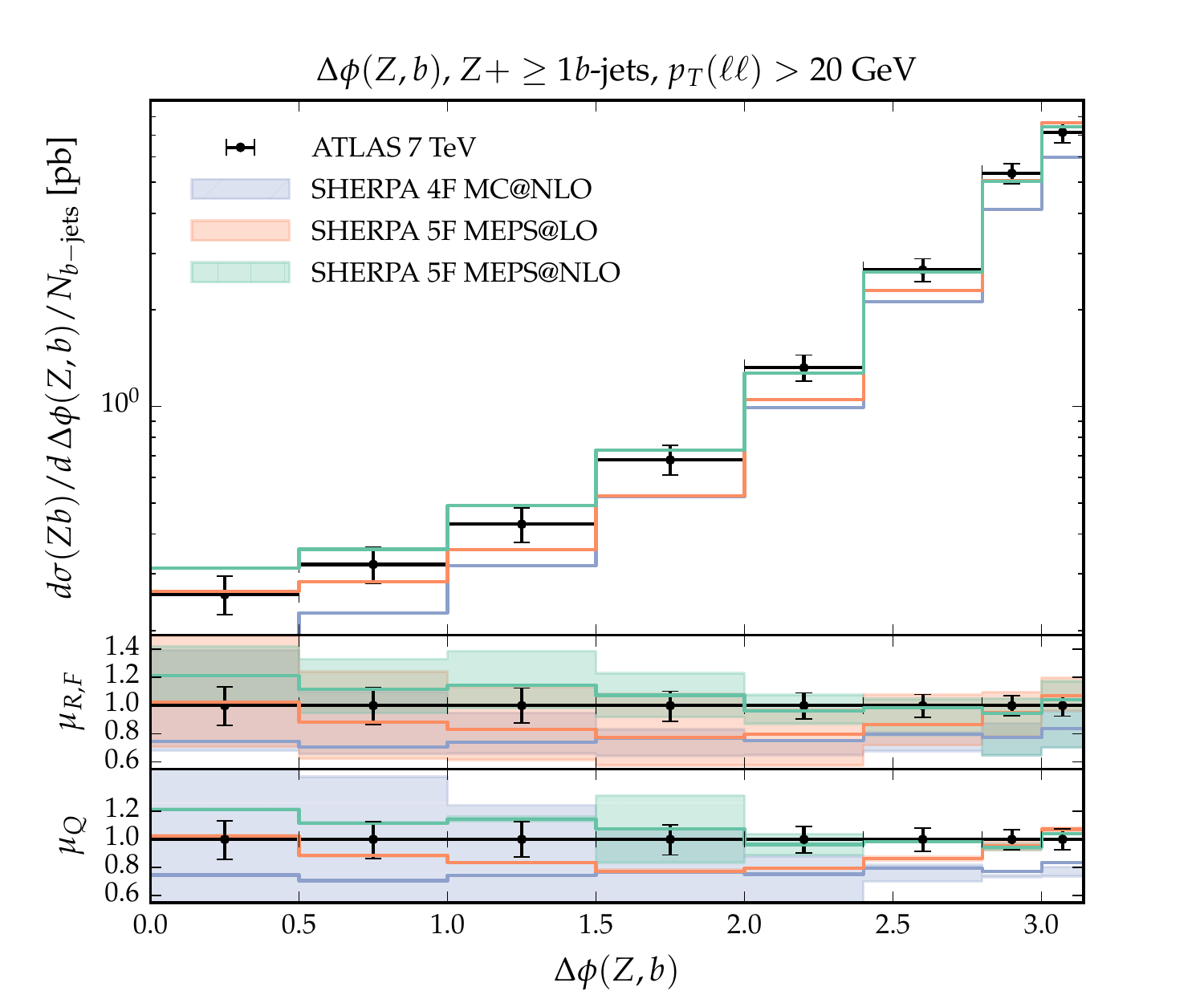}
  \caption{Transverse-momentum distribution of the Z boson (left) and the 
    azimuthal separation between the $Z$ boson and the $b$-jets (right) 
    in events with at least one $b$-jet. For the $\Delta\phi(Z,b)$ measurement 
    the additional constraint $p_{T,ll}>20\;{\rm GeV}$ is imposed.
    Data taken from Ref.~\cite{Aad:2014dvb}. 
    \label{fig:1bzpt}}
}
\end{figure}

This pattern is repeated in Fig.~\ref{fig:1bzpt}, where we show
the differential $\sigma(Zb)$ cross section with respect to the dilepton
transverse momentum and, rescaled to $1/N_{b-{\rm jets}}$, as a
function of the azimuthal separation between the reconstructed $Z$ boson
and the $b$-jets.  Again, both distributions are very well modelled by both
5F calculations. The 4F \MCatNLO prediction again underestimates data by
a largely flat 20-25\%.

\begin{figure}[!htb]
\centering{
  \includegraphics[width=0.45\textwidth]{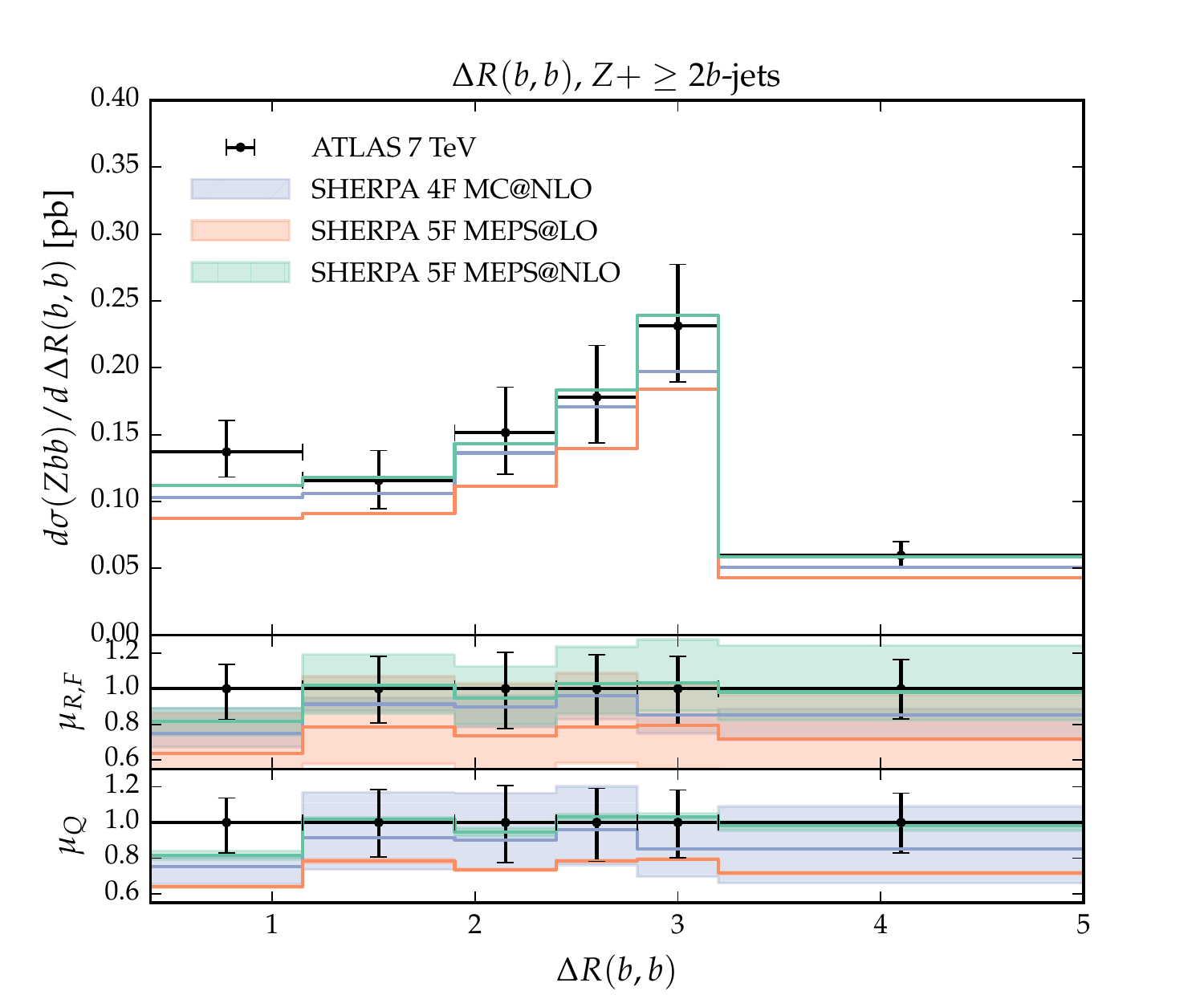}
  \includegraphics[width=0.45\textwidth]{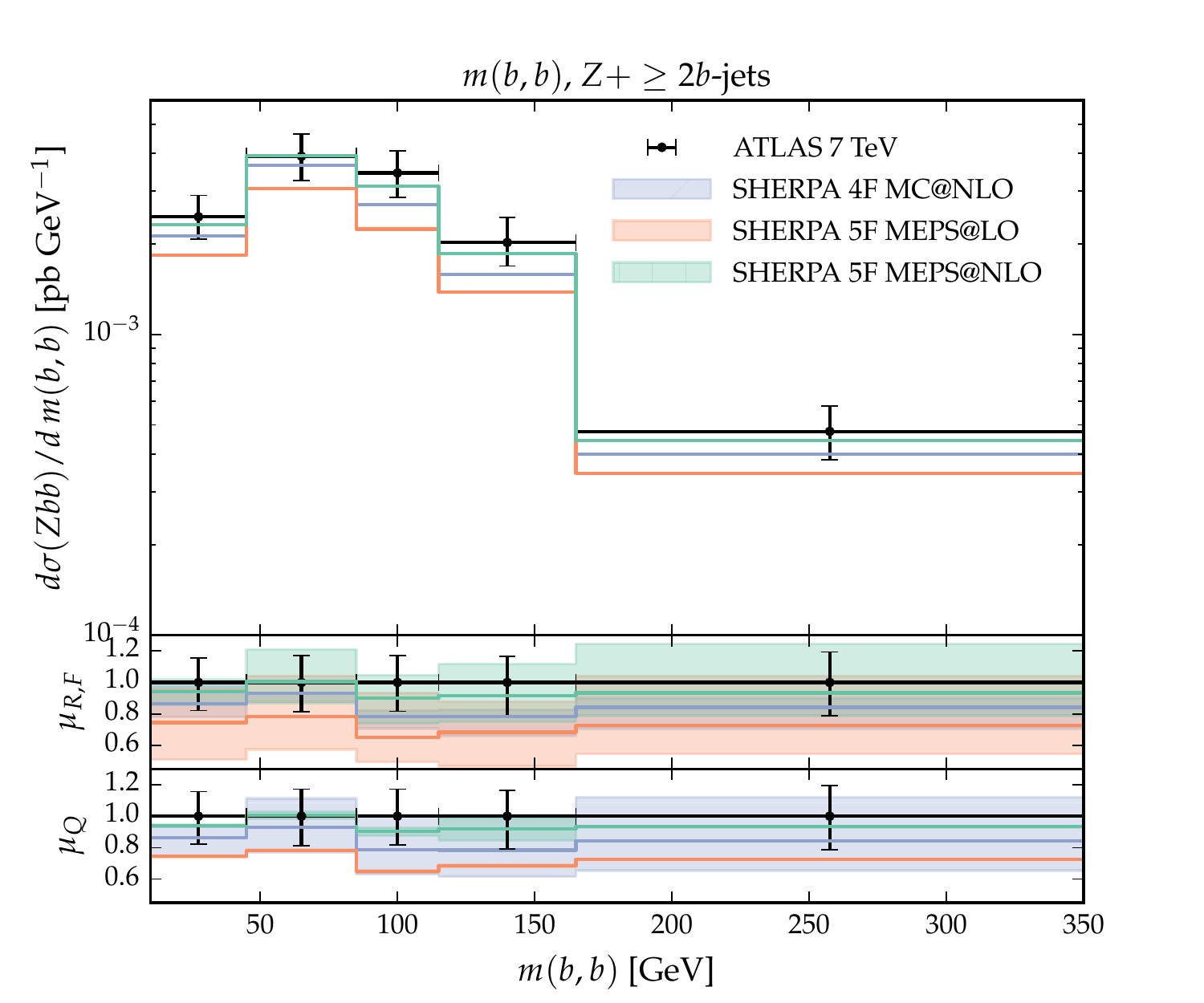}
  \caption{The $\Delta R$ separation (left) and invariant-mass distribution
    (right) for the leading two $b$-jets. Data taken from
    Ref.~\cite{Aad:2014dvb}.
    \label{fig:2b}}
}
\end{figure}

Moving on to final states exhibiting at least two identified $b$-jets,
the role of the 5F~LO and 4F~NLO predictions are somewhat reversed:
As can be inferred from Fig.~\ref{fig:xstot}, the 4F and 5F~NLO samples
provide good estimates for the inclusive $Zbb$ cross section, while the
5F~LO calculation undershoots data by about $20$\%.  In Fig.~\ref{fig:2b}
the $\Delta R$ separation of the two highest transverse-momentum $b$-jets
along with their invariant-mass distribution is presented. Both the 4F and
the 5F approaches yield a good description of the shape of the distributions. 
It is worth stressing that this includes the regions of low invariant mass 
and low $\Delta R$, corresponding to a pair of rather collinear $b$-jets.
This is a region that is usually riddled by potentially large logarithms,
where the parton shower starts taking effect.  Note that in the comparison
presented in~\cite{Aad:2014dvb} this region showed some disagreement between
data and other theoretical predictions based on NLO QCD (dressed with parton
showers). 

\begin{figure}[!htb]
\centering{
  \includegraphics[width=0.45\textwidth]{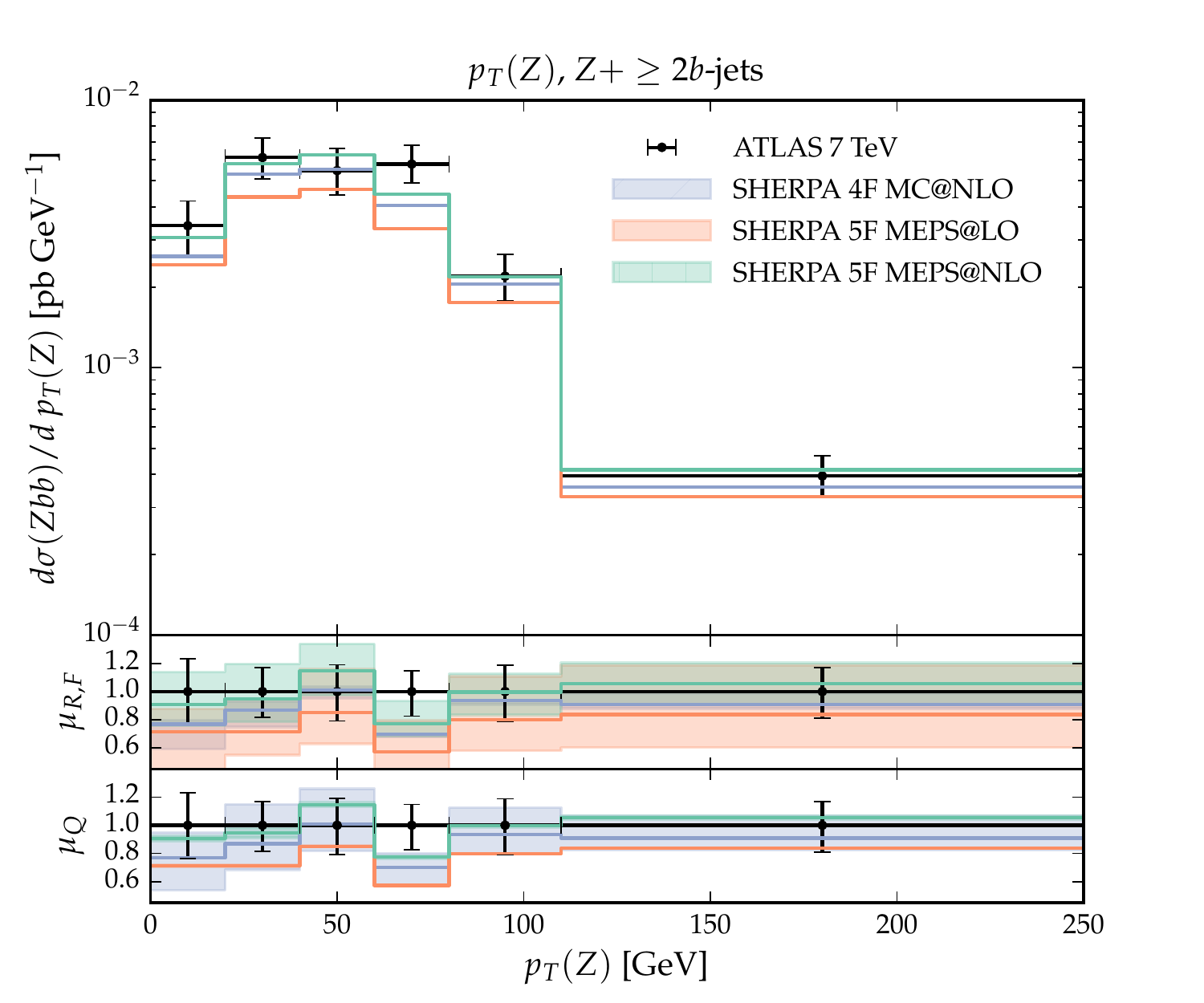}
  \caption{Transverse-momentum distribution of the dilepton system for events
    with at least two $b$-jets. Comparison against various calculational 
    schemes.
    Data taken from Ref.~\cite{Aad:2014dvb}.
    \label{fig:2bzpt}}
}
\end{figure}

In Fig.~\ref{fig:2bzpt} the resulting transverse-momentum distribution of 
the dilepton system when selecting for events with at least two associated 
$b$-jets is shown. The shape of the data is very well reproduced by the
4F~\MCatNLO and 5F~\MEPSatNLO samples. Also the 5F~\MEPSatLO prediction
describes the data well despite of the overall rate being $20$\% lower than
observed in data.

\begin{figure}[!htb]
\centering{
  \includegraphics[width=0.45\textwidth]{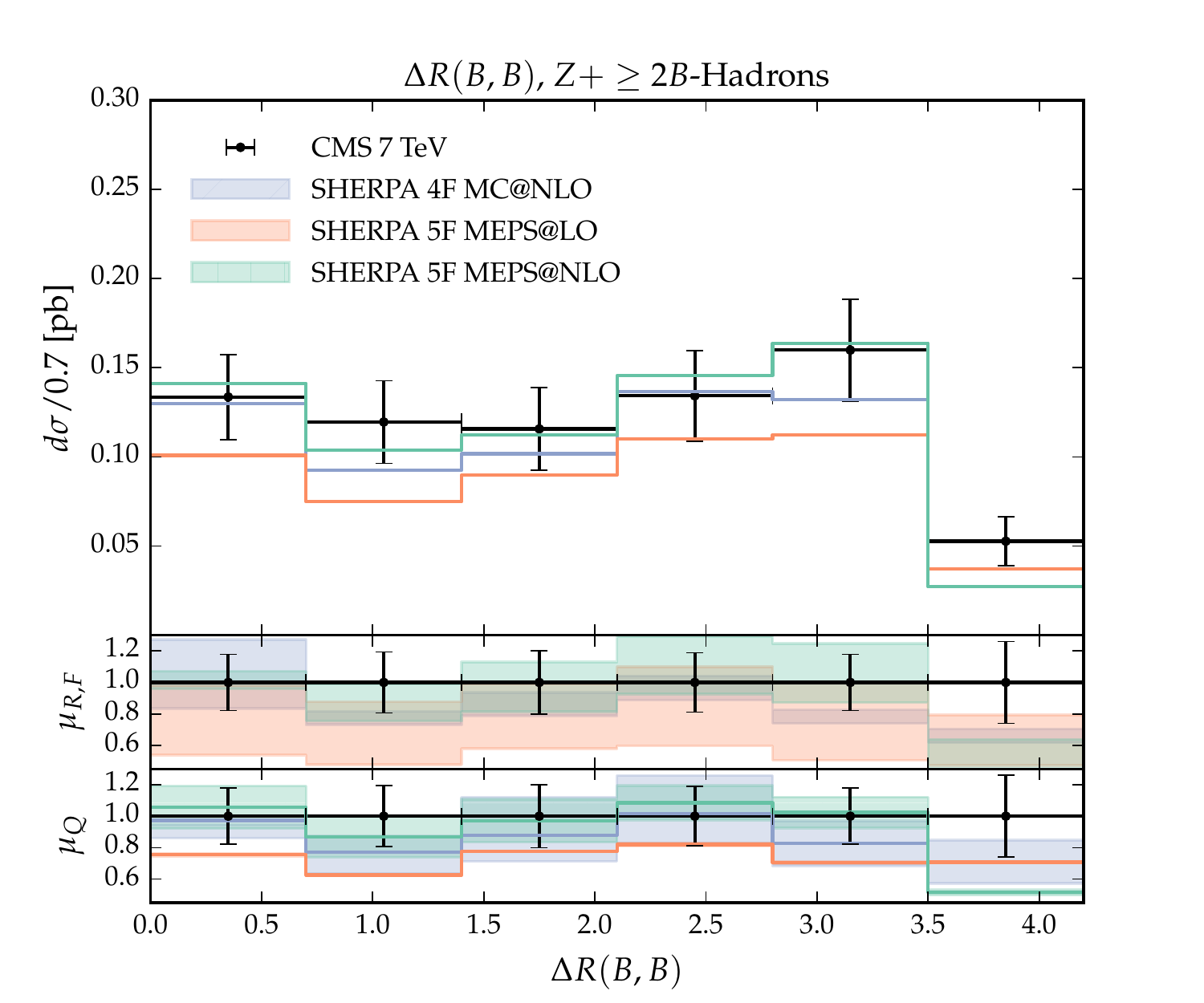}
  \includegraphics[width=0.45\textwidth]{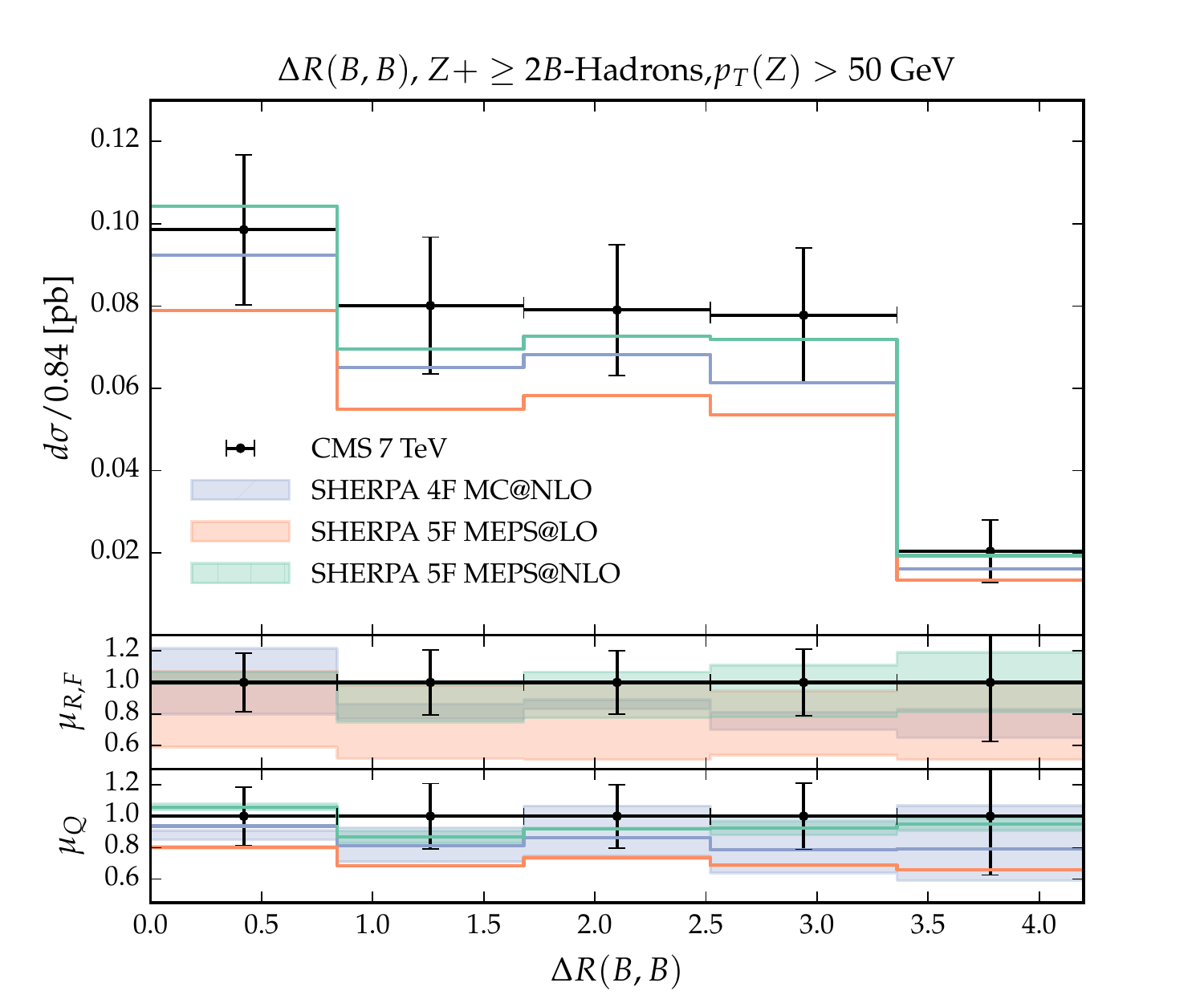}
  \caption{ $\Delta R_{BB}$ distribution for two selections of the transverse 
    momentum of the $Z$ boson. Data taken from Ref.~\cite{Chatrchyan:2013zja}.
    \label{fig:2bdR_CMS}}
}
\end{figure}

\begin{figure}[!htb]
\centering{
  \includegraphics[width=0.45\textwidth]{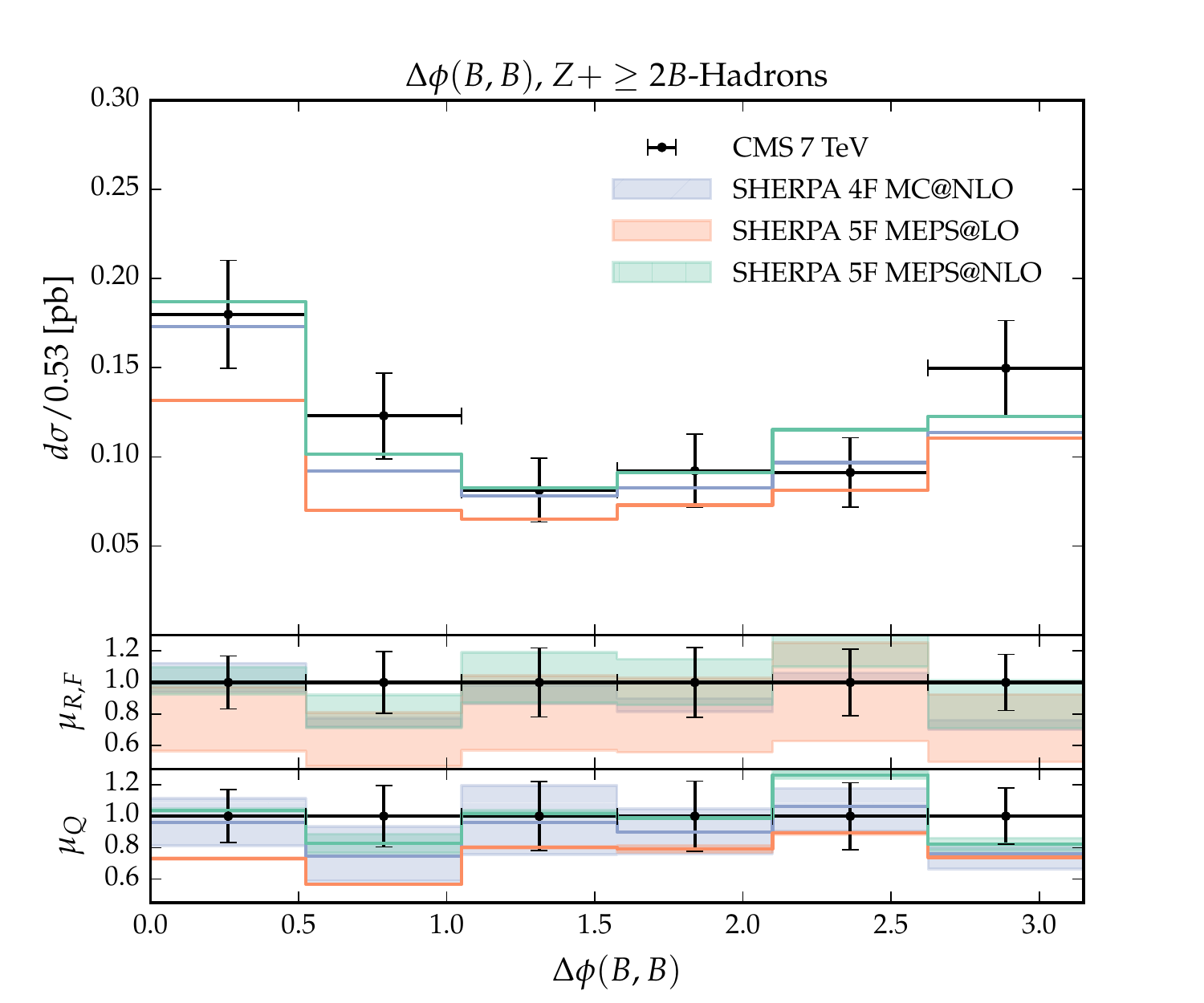}
  \includegraphics[width=0.45\textwidth]{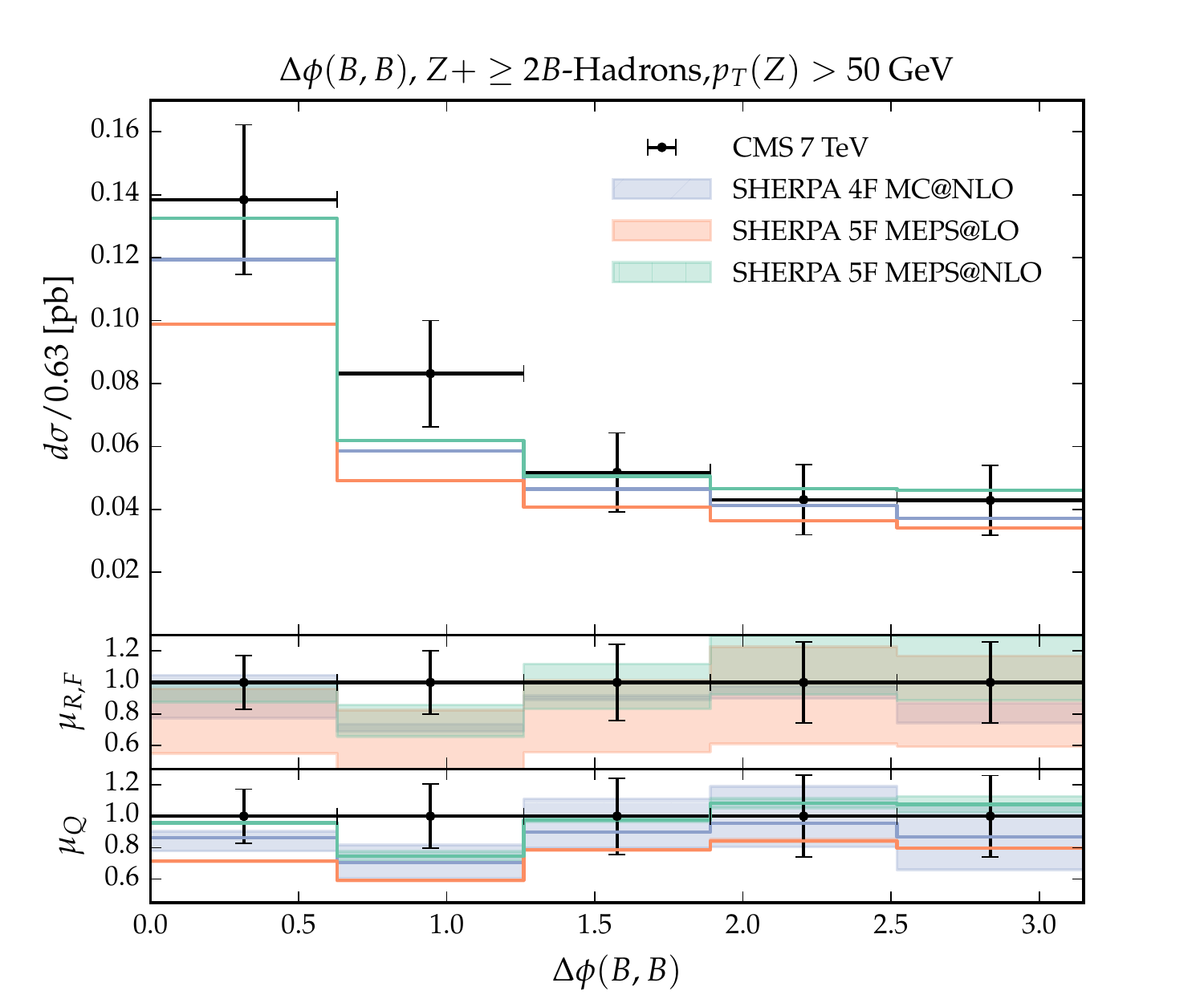}
  \caption{ $\Delta \phi_{BB}$ distribution for two selections of the transverse 
    momentum of the $Z$ boson. Data taken from Ref.~\cite{Chatrchyan:2013zja}.
    \label{fig:2bdPhi_CMS}}
}
\end{figure}

The measurements presented by the CMS collaboration in
Ref.~\cite{Chatrchyan:2013zja} focus on angular correlations between
$b$-hadrons rather than $b$-jets.  Two selections with respect to the
dilepton transverse momentum have been considered, a sample requiring
$p_T(Z)>50\;{\rm GeV}$ and an inclusive one considering the whole range
of $p_T(Z)$.  The $\Delta R$ and $\Delta \phi$ separation of the
$b$-hadrons obviously prove to be most sensitive to the theoretical
modelling of the $b$-hadron production mechanism and the interplay of
the fixed-order components and the parton showers.  They are presented in
Figs.~\ref{fig:2bdR_CMS} and \ref{fig:2bdPhi_CMS}.  In general, a good
agreement in the shapes of simulation results and data is found, with
the same pattern of total cross sections as before: the 5F~\MEPSatNLO
sample describes data very well, while the 4F~\MCatNLO results tend to
be a little bit, about 10\%, below data, with data and theory 
uncertainty bands well overlapping, while the central values of the 
5F~\MEPSatLO results undershoot data by typically 20-25\%.

Overall it can be concluded that the 5F \MEPSatNLO calculation yields the best 
description of the existing measurements, regarding both the production rates 
{\em and} shapes. The 4F \MCatNLO and 5F \MEPSatLO schemes succesfully model 
the shape of the differential distributions but consistently underestimate the 
production rates.

\section{Bottom-jet associated Higgs-boson production}
\label{sec:hbb}
In this section we present predictions for $b$-jet(s) associated production
of the Standard-Model Higgs boson in $pp$ collisions at the $13$ TeV \LHC
obtained in the four-- and five--flavour schemes. As standard when dealing with this
process, we do not include contributions from the gluon-fusion channel.
However, in the 4F \MCatNLO we do include terms proportional to the top-quark 
Yukawa coupling, contributing to order $y_by_t$ as an interference effect at 
NLO QCD~\cite{Dittmaier:2003ej,Dawson:2003kb,deFlorian:2016spz}.  
Although associated $Z+$ $b$-jet(s) production serves as a good proxy for the 
Higgs-boson case, there are important differences between both processes, mainly 
due to the different impact of initial-state light quarks, which couple to $Z$ 
bosons but not to the Higgs boson.  

As before, QCD jets are defined through the anti-$k_t$ algorithm using a
radius parameter of $R=0.4$, a minimal transverse momentum $p_{T,j}>25$~GeV,
and a rapidity cut of $|y_{j}|<2.5$.  In this case, we consider results that are at
the parton level only, disregarding hadronisation and underlying-event
effects, which may blur the picture.  We consider single $b$-tagged jets only,
thus excluding jets with intra-jet $g\rightarrow b\bar{b}$ splittings from
the parton shower which would be the same for all flavour schemes we
investigate.  As for $Z$-boson production, we separate the event samples
into categories with at least one $b$-jet, i.e.\ $H+\geq 1 b$-jet events,
and at least two tagged $b$-jets, i.e.\ $H+\geq 2 b$-jets events. 

\begin{table}[!hbt]
  \centering
  \caption{$13$~TeV total cross sections and the corresponding 
    $\mu_{F/R}$ and $\mu_Q$ uncertainties for $H + \geq 1 b$ and $H + \geq 2 b$s. 
  }\label{tab:hbbxs}
  \begin{tabular}[\linewidth]{lc|c}
    \toprule
    \LHC 13~TeV & $H + \geq 1 b$-jets [fb] & $H + \geq 2 b$-jets [fb]\\
    \midrule
    $\sigma_{\text{\MCatNLO}}^{4F}$   &
    $45.2^{+15.5\%}_{-18.4\%}$  &
    $4.5^{+25.1\%}_{-26.3\%}$\\
    $\sigma_{\text{\MEPSatLO}}^{5F}$  &
    $79.3^{+34.0\%}_{-25.4\%}$  &
    $3.8^{+34.3\%}_{-30.3\%}$\\
    $\sigma_{\text{\MEPSatNLO}}^{5F}$ &
    $110.5^{+14.2\%}_{-16.0\%}$    &
    $6.9^{+27.3\%}_{-27.1\%}$\\
    \bottomrule
  \end{tabular}
\end{table}

\begin{figure}[!htb]
\centering
  \includegraphics[width=0.4\textwidth]{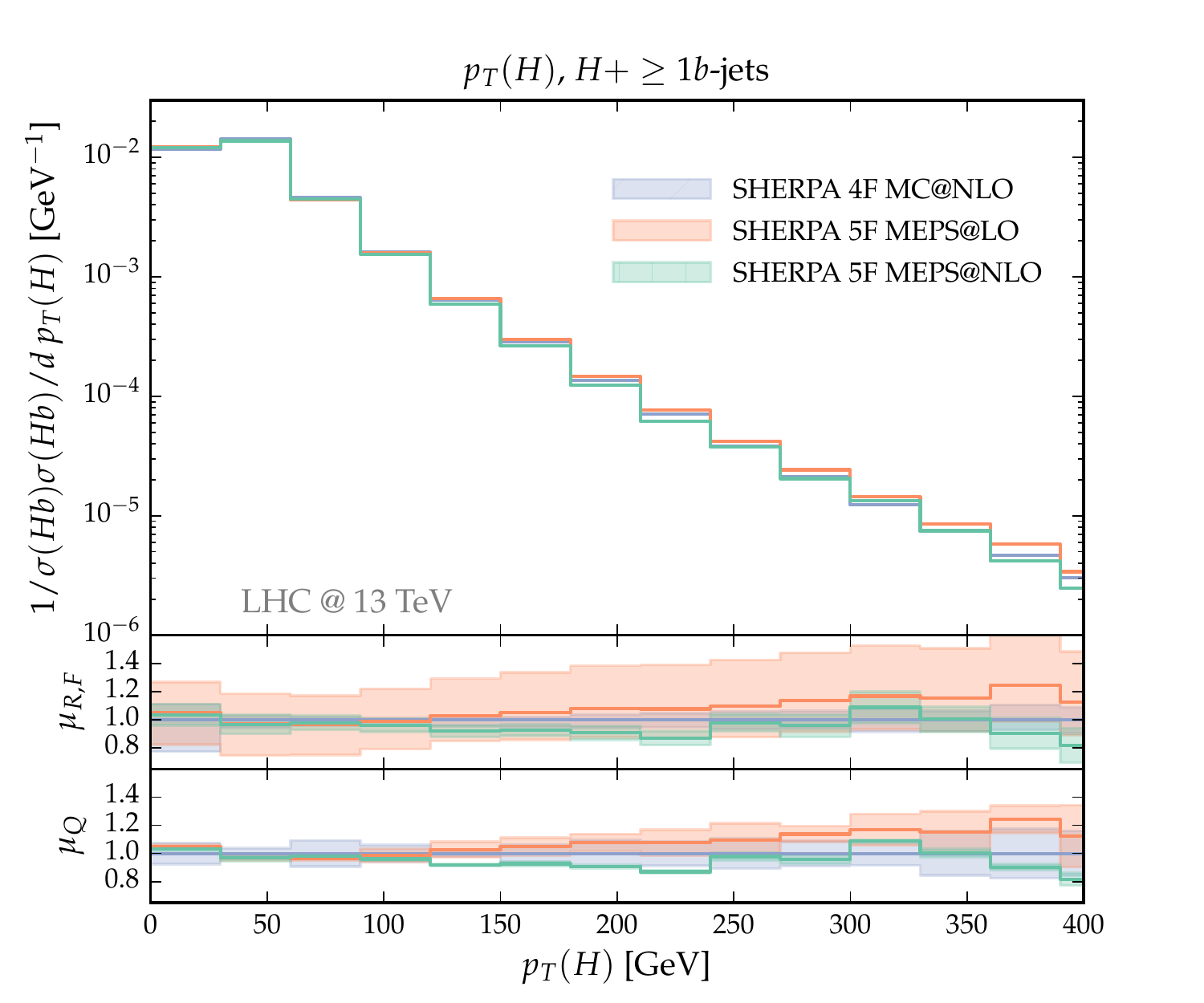}
  \includegraphics[width=0.4\textwidth]{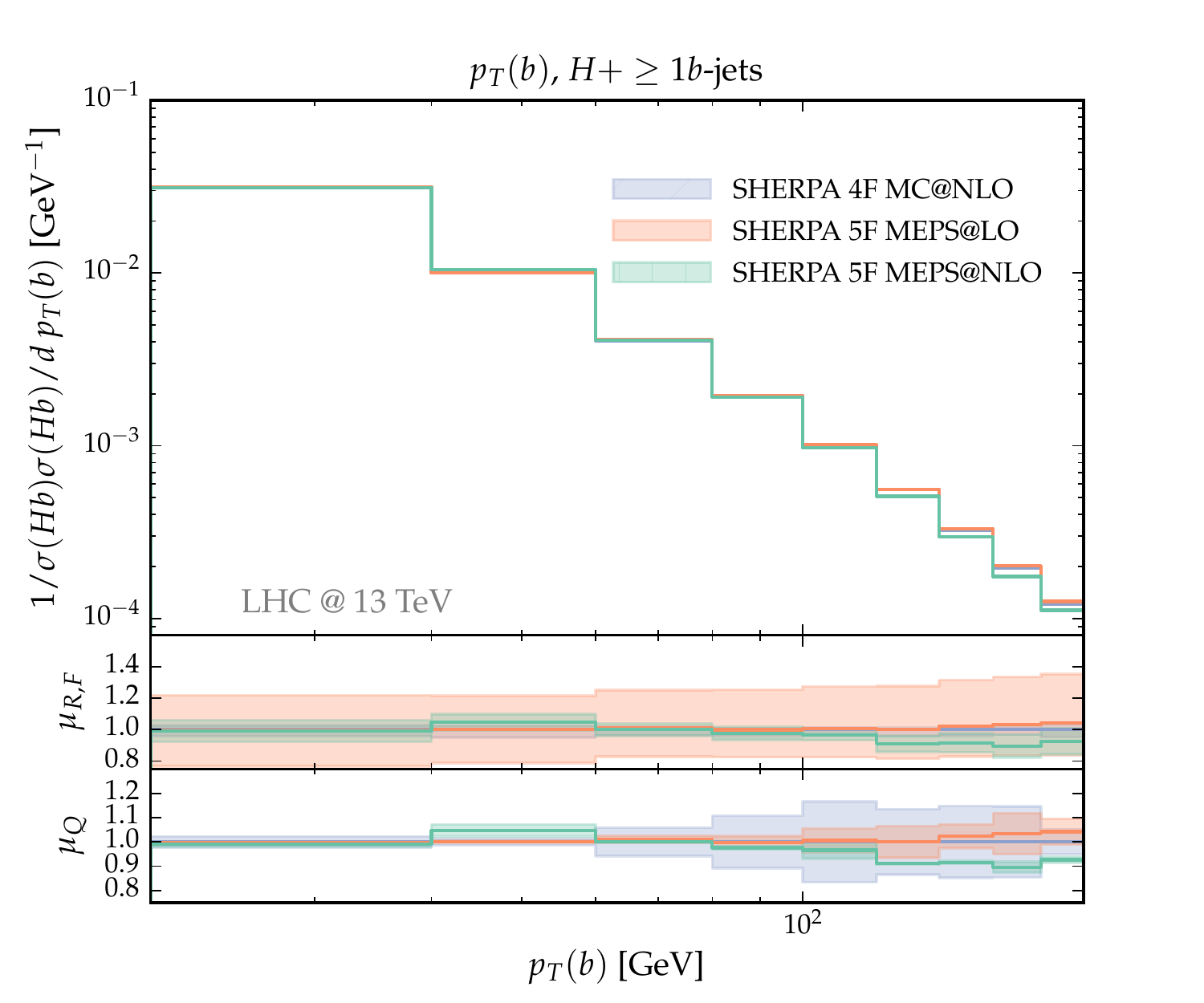}
  \caption{Predictions for the transverse-momentum distribution of the Higgs
    boson (left panel) and the leading $b$-jet (right panel) in inclusive
    $H+b$-jet production at the $13$ TeV \LHC.}\label{fig:1bh}
\end{figure}

\begin{figure}[!htb]
\centering
  \includegraphics[width=0.5\textwidth]{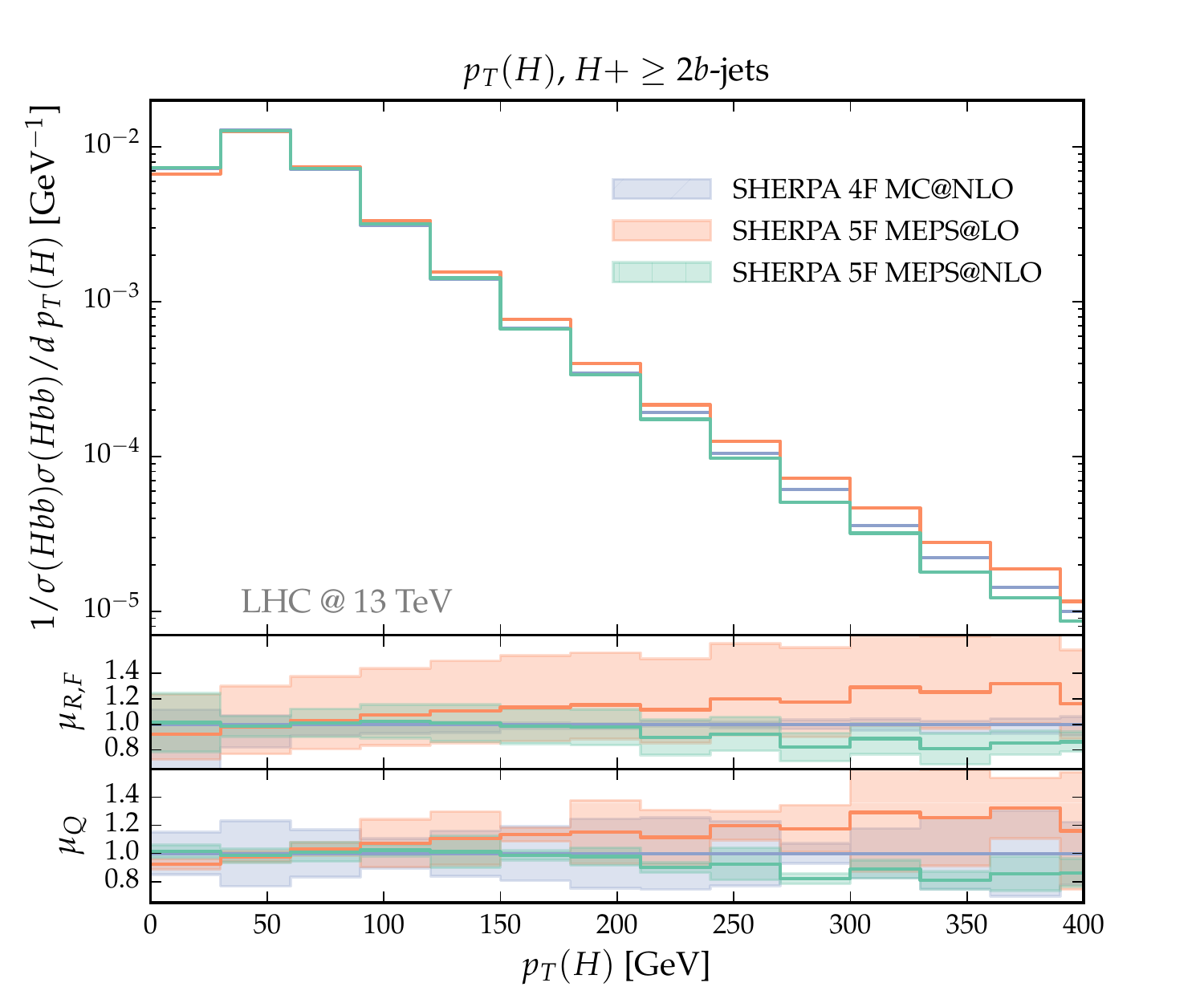}
  \caption{The transverse-momentum distribution of the Higgs boson in
    inclusive $H+2b$-jets production at the $13$ TeV \LHC.}\label{fig:1bhpt}
\end{figure}

\begin{figure}[!htb]
\centering
  \includegraphics[width=0.4\textwidth]{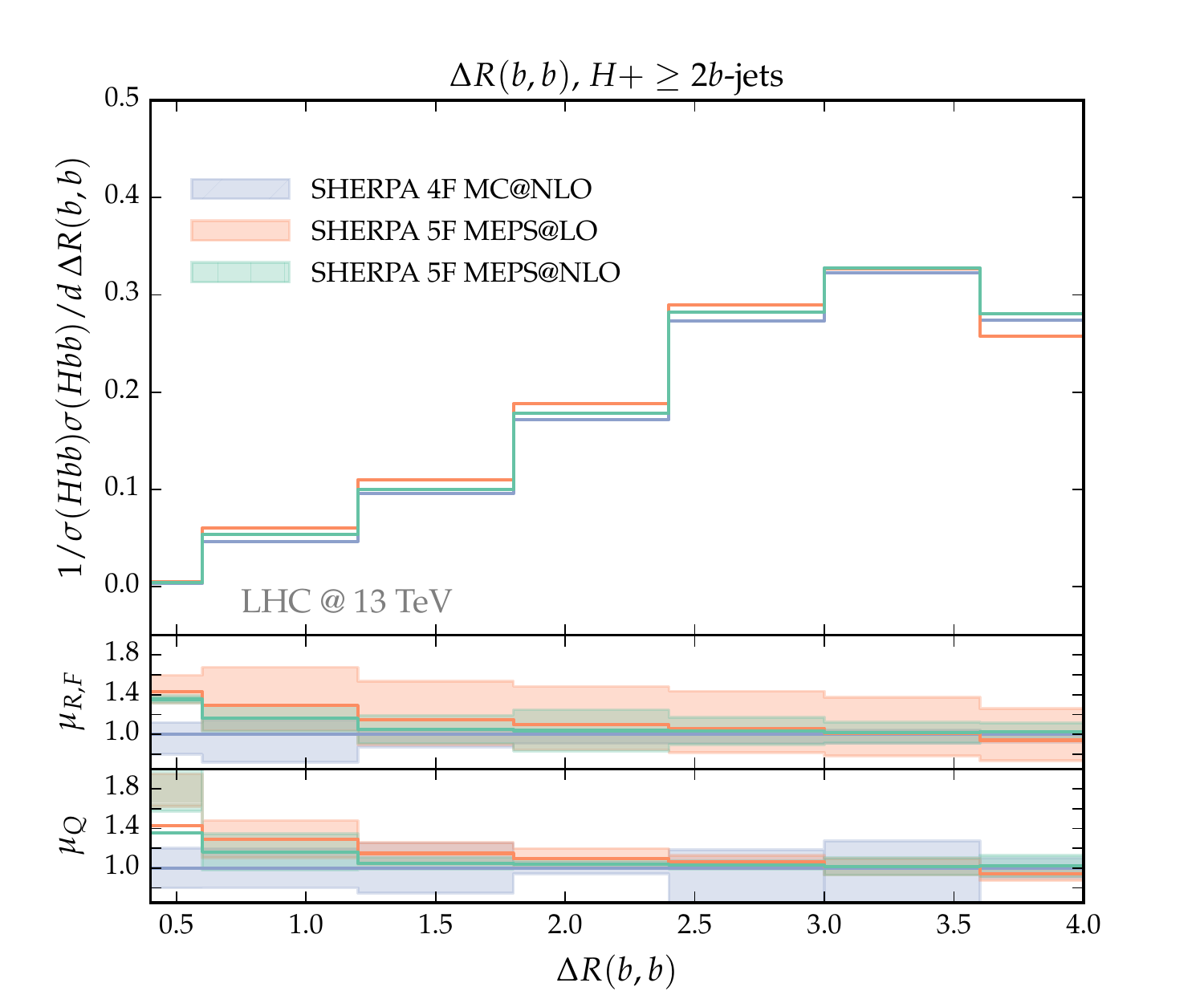}
  \includegraphics[width=0.4\textwidth]{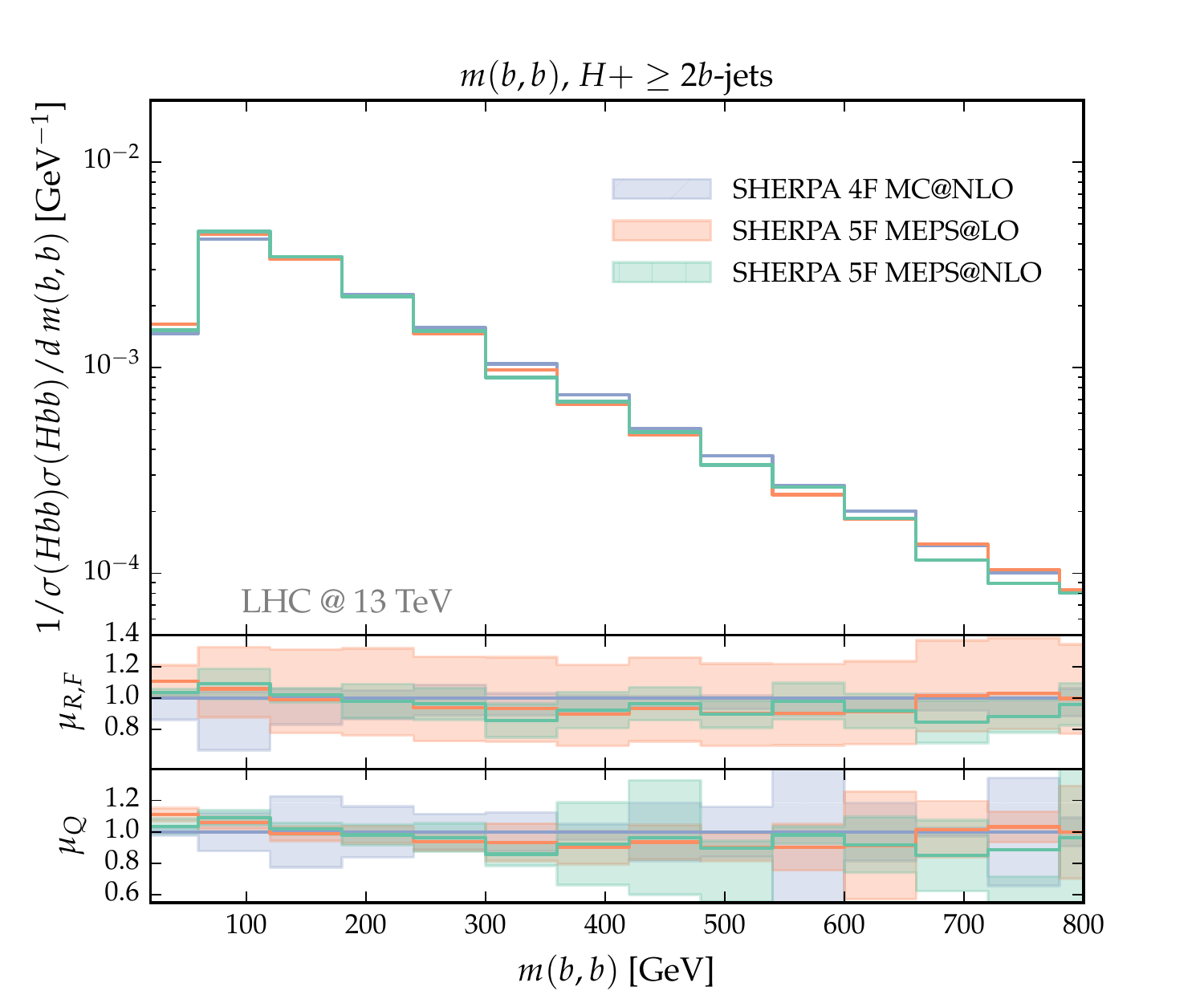}
  \caption{Predictions for the $\Delta R$ separation of the two leading
    $b$-jets (left panel) and their invariant-mass distribution
    (right panel) in inclusive $H+2b$-jets production at the 
    $13$ TeV \LHC.}\label{fig:2bh}
\end{figure}
In Tab.~\ref{tab:hbbxs} cross sections for the three calculations are
reported.  Historically, inclusive results have largely disagreed between the
4F and the 5F scheme.  This feature is observed for the case at hand, too,
and especially so for the case of one tagged $b$-jet.  There the
4F \MCatNLO prediction is smaller than the 5F results by factors of about
$1.75$ (5F LO) and of $2.44$ (5F NLO).  The relative differences are
reduced when a second tagged $b$-jet is demanded.  In this case we find that
the 4F result lies between the two 5F results, about 20\% higher than the
LO predictions, and a factor of about 1.5 lower than the 5F NLO predictions.
In both cases, inclusive $H+b$ and $H+bb$ production, the uncertainty bands
of the two 5F predictions, corresponding to 7-point $\mu_{R/F}$ variations 
and $\mu_Q$ variations by a factor of two up and down, do overlap.  While for 
the two $b$-jet final states this includes the 4F result, for the one $b$-jet 
case the 4F result is not compatible with the 5F predictions, taking into
account the considered scale uncertainties. It is worth noting that a milder 
form of this relative scaling of the cross sections was already observed
in the $Z$ case.

In the case of the total inclusive cross section, this very large difference
can be mitigated by including higher-order corrections, on the one hand, and a
better assessment of which choice of the unphysical scales yields the better
agreement~\cite{Frederix:2011qg,Wiesemann:2014ioa,deFlorian:2016spz,Lim:2016wjo}.
However, only a recent effort to match the two schemes~\cite{Forte:2015hba,
  Forte:2016sja,Bonvini:2015pxa, Bonvini:2016fgf}
has clearly assessed the relative importance of mass corrections
(appearing in the 4F scheme) and large log resummation (as achieved in a 5F
scheme). In particular it has been found that the difference between these two
schemes is mostly given by the resummation of large logarithms, thus suggesting that
for an inclusive enough calculation either a 5F scheme or a matched scheme
should be employed.  This is the same situation that one faces, albeit
milder, in the $Z$ case, where, in terms of normalisation the 5F scheme
performed better in all cases and especially in inclusive calculations.  We
therefore recommend that in terms of overall normalisation, the 5F \MEPSatNLO 
scheme should be used to obtain reliable predictions.

Let us now turn to the discussion of the relative differences in the
shapes of characteristic and important distributions.  To better appreciate
shape differences, all differential distribution are normalised to the
respective cross section, i.e.\ the inclusive rates $\sigma(Hb)$ and
$\sigma(Hbb)$.  In all cases we obtain agreement at the 15\%-level or better 
between the 5F~\MEPSatNLO and 4F~\MCatNLO samples, the only exception, not
surprisingly, being the region of phase space where the two $b$'s come
close to each other and resummation effects start playing a role.
Typically, the 5F~\MEPSatLO predictions are also in fair agreement with the
other two results, however, they exhibit a tendency for harder tails in
the $p_T$ distributions, mainly in the inclusive Higgs-boson $p_T$ and
in the transverse momentum of the second $b$ jet.  

Starting with Fig.~\ref{fig:1bh}, the transverse-momentum distributions of the 
Higgs boson and the leading $b$-jet in the case of at least one $b$-jet tagged is
displayed. Similarly to the $Z$ example, this is the region where one would
expect the 5F scheme to perform better. However, again similarly to the $Z$
case, the three schemes largely agree in terms of shapes, being well within
scale uncertainties. Notably, this turns out to be particularly true for
the low ($\sim 20$--$100$~\UGeV) $p_T$ region where one could have expected
deviations to be the largest.

In Figs.~\ref{fig:1bhpt} and \ref{fig:2bh} we present differential
distributions for the selection of events with at least two tagged
$b$-jets. While Fig.~\ref{fig:1bhpt} shows the resulting Higgs-boson
transverse-momentum distribution, Fig.~\ref{fig:2bh} compiles results for
the $\Delta R$ separation of the two leading $b$-jets and their invariant-mass
distribution.  For such two $b$-jets observables the 4F scheme is expected
to work best, especially when the two $b$ are well separated to suppress
potentially large logarithms.  However, in agreement with the $Z$-boson
case, no significant differences between the various scheme arise when
taking into account $\mu_{R/F}$ and $\mu_Q$ scale-variation uncertainties.  
Once again the region of low $p_T$ in Fig.~\ref{fig:1bhpt} and the region of 
low $m(b,b)$ in Fig.~(\ref{fig:2bh}) show excellent agreement amongst the various
descriptions.  As anticipated, larger differences can be seen between the
two 5FS and the 4F \MCatNLO calculations, in the very low $\Delta R(b,b)$ and
$m(b,b)$ regions, Fig.~(\ref{fig:2bh}), where the two $b$-jets become
collinear.  This feature is however most likely due to the fact that we are
dealing with partonic $b$-jets as opposed to {\em hadronic} ones. Taking
as a reference the $Z$-boson case once again, in fact, where this difference
is not present at all, suggests that a realistic simulation, that
accounts for hadronisation effects, should largely suppress this difference.

\section{Conclusions}

Simulations for the associated production of a $Z$- or a Higgs-boson with a $b\bar{b}$
pair have always proven to require careful thinking in including or neglecting
$b$-quark mass effects.  In this work, a detailed comparison between the 4F and the
5F schemes implemented in \Sherpa has been presented.

Firstly, the results for production of a $Z$ boson with $b$-jets has been compared
with both \ATLAS and \CMS data.  We find that all schemes largely agree in
the shapes of relevant observables.  Major differences however appear in 
the overall normalisation of the various samples, with the 5F \MEPSatNLO
prediction being the one proving the best agreement with data.

We used this as a guide to study the $b$-associated production of Standard-Model 
Higgs bosons.  Due to the different impact of the initial-state $b$-quarks,
this process enhances the quantitative differences between the different
approaches, and in particular the production cross sections, while it still maintains
the qualitative scaling behaviour.  This qualitative similarity is fortified
by the good agreement of the calculations in the shapes of sensitive
observables.

We thus conclude that in order to obtain reliable predictions, at the \LHC,
for the production of a Higgs boson with $b$-jets, the use of a 5F
\MEPSatNLO set--up is the most advisable.  A second, more phenomenologically
driven, option could be to use a 4FS, \MCatNLO accurate, prediction,
normalised by the 5F \MEPSatNLO total cross section. This is particularly
relevant given that the 4FS calculation is by far the most efficient one.

To further improve our theoretical predictions, we plan to generalise the 
treatment of finite-mass effects in the 5FS by allowing for massive initial-state
quarks in the matrix elements and the parton showers. Besides using fully mass-dependent
matrix elements also at NLO, this requires a generalisation of the implementation of 
the NLO subtraction formalism, along the lines of \cite{Dittmaier:1999mb},  as well as 
the inclusion of mass effects in the initial-state parton-shower splitting functions.   

\section*{Acknowledgements}
We want to thank our colleagues from the \Sherpa collaboration 
for fruitful discussions and technical support. We acknowledge 
financial support from the EU research networks funded by the Research
Executive  Agency (REA) of the European Union under Grant Agreements 
PITN-GA2012-316704 (``HiggsTools'') and PITN-GA-2012-315877 (``MCnetITN''),
by the ERC Advanced Grant MC@NNLO (340983), and from BMBF under contracts
05H12MG5 and 05H15MGCAA,

\bibliographystyle{amsunsrt_modp}
\bibliography{bbb}
\end{document}